\pdfoutput=1
\documentclass[a4paper,11pt]{article}
\usepackage[utf8]{inputenc}

\usepackage[table]{xcolor}
\usepackage{jcappub}
\usepackage{graphicx}
\usepackage{psfrag,fancyhdr,epsfig}
\usepackage{hyperref}
\hypersetup{colorlinks,bookmarksopen,bookmarksnumbered,citecolor=magenta,linkcolor=blue,pdfstartview=FitH,urlcolor=blue}

\usepackage{amsmath}
\usepackage{array}
\usepackage{graphicx}
\usepackage{color}
\usepackage{tensor}
\usepackage{xcolor}
\usepackage{orcidlink}

\newcommand{\be}{\begin{equation}}
\newcommand{\ee}{\end{equation}}
\newcommand{\bear}{\begin{array}}
\newcommand{\eear}{\end{array}}
\newcommand{\ba}{\begin{eqnarray}}
\newcommand{\ea}{\end{eqnarray}}

\def\a{\alpha}
\def\b{\beta}

\def\d{\delta}

\def\eps{\epsilon}
\def\g{\gamma}

\def\l{\lambda}
\def\m{\mu}
\def\n{\nu}

\def\r{\rho}
\def\s{\sigma}

\def\tns{\tensor}
\def\cR{\mathcal{R}}
\def\cC{\mathcal{C}}

\newcommand{\GeV}{\; \mathrm{GeV}}
\newcommand{\TeV}{\; \mathrm{TeV}}

\usepackage[most]{tcolorbox}

\tcbset{colback=yellow!10!white, colframe=red!50!black, 
        highlight math style= {enhanced, %<-- needed for the ’remember’ options
            colframe=red,colback=red!10!white,boxsep=0pt}
        }

        \usepackage{empheq}
\newtcbox{\mymath}[1][]{%
    nobeforeafter, math upper, tcbox raise base,
    enhanced, colframe=blue!30!black,
    colback=blue!30, boxrule=1pt,
    #1}

\title{Inflation and reheating in quadratic metric-affine gravity with derivative couplings}


\author[a]{Ioannis D.~Gialamas\orcidlink{0000-0002-2957-5276}}
\author[b]{, Theodoros Katsoulas\orcidlink{0000-0003-4103-7937}}
\author[b]{ and Kyriakos Tamvakis\orcidlink{0009-0007-7953-9816}}

\emailAdd{ioannis.gialamas@kbfi.ee}
\emailAdd{th.katsoulas@uoi.gr} 
\emailAdd{tamvakis@uoi.gr}

\affiliation[a]{Laboratory of High Energy and Computational Physics, 
National Institute of Chemical Physics and Biophysics, R{\"a}vala pst.~10, Tallinn, 10143, Estonia}
\affiliation[b]{Physics Department, University of Ioannina, 45110, Ioannina, Greece}

\abstract{Within the framework of metric-affine theories of gravity, where both the metric and connection are treated as independent variables, we consider actions quadratic in the Ricci scalar curvature coupled non-minimally to a scalar field through derivative couplings. Our analysis delves into the inflationary predictions, revealing their consistency with the latest observational constraints across a wide range of parameters. This compatibility permits adjustments such as an increase in the spectral index and a reduction in the tensor-to-scalar ratio. While we do not propose a specific reheating mechanism, our analysis demonstrates that within the quadratic model of inflation, the maximum reheating temperature can reach $\sim 3\times10^{15} \GeV$.}

\begin{document}

\maketitle

\section{Introduction}
\label{introduction}
 
Cosmological inflation~\cite{Kazanas:1980tx, Sato:1980yn, Guth:1980zm, Linde:1981mu} offers a natural explanation of how quantum fluctuations of gravitational and matter fields can be promoted to the cosmological perturbations~\cite{Starobinsky:1979ty, Mukhanov:1981xt, Hawking:1982cz, Starobinsky:1982ee, Guth:1982ec, Bardeen:1983qw} that are the origin of large scale structure of the Universe. The standard way of realizing inflation is through the vacuum energy of a scalar degree of freedom (the inflaton), either introduced as a fundamental field or as part of gravity itself. In both cases non-minimal couplings of the inflaton to curvature are expected to be present as corrections to the classical general relativity (GR) action arising from the quantum interactions of gravitating matter fields. Such are couplings of the scalar field to the Ricci scalar curvature $\sim f(\phi){R}$~\cite{Bezrukov:2007ep} or derivative couplings to the Ricci tensor $\sim (\partial_{ \mu}\phi\partial_{ \nu}\phi)R^{ \mu\nu}$~\cite{Amendola:1993uh, Capozziello:1999uwa, Capozziello:1999xt, Germani:2010gm, Tsujikawa:2012mk,Kamada:2012se, Sadjadi:2012zp, Koutsoumbas:2013boa,Ema:2015oaa, Gumjudpai:2015vio, Zhu:2015lry, Sheikhahmadi:2016wyz,Dalianis:2016wpu,Harko:2016xip, Tumurtushaa:2019bmc, Fu:2019ttf,Dalianis:2019vit, Sato:2020ghj, Karydas:2021wmx}. Quadratic terms of the curvature invariants~\cite{Stelle:1976gc} are also expected to modify the classical action, a particular case being that of a quadratic Ricci scalar term $\sim R^2$ giving rise to the Starobinsky model of inflation~\cite{Starobinsky1980}.

The standard metric formulation of GR, where the connection is given by the Levi-Civita relation in terms of the metric, is known to be equivalent to the metric-affine formulation, where the connection is an independent variable. Nevertheless, this equivalence ceases to be true if we depart from the Einstein-Hilbert form of the action and consider modifications such as non-minimal couplings of scalar fields to the curvature or quadratic curvature terms. In the general framework of metric-affine theories of gravity, while general covariance is preserved, the connection is promoted to a variable independent of the metric. The difference of the independent connection and the Levi-Civita one is the distortion tensor, which in cases that it can be integrated out, leads to a resulting effective metric theory with or without additional dynamical degrees of freedom. A characteristic example is the case of the metric-affine (Palatini) version of the ${\cal{R}}^2$ model~\cite{Meng:2004yf} which, in contrast to the analogous metric model, does not predict any propagating scalar mode, although, when coupled to a scalar field, gives rise to a characteristic inflationary plateau independently of the scalar self interaction potential. This general feature of metric-affine ${\cal{R}}^2$ models~\cite{Meng:2004yf, Borunda:2008kf, Bombacigno:2018tyw, Enckell:2018hmo, Iosifidis:2018zjj, Antoniadis:2018ywb, Antoniadis:2018yfq, Tenkanen:2019jiq, Edery:2019txq, Giovannini:2019mgk, Gialamas:2019nly,  Lloyd-Stubbs:2020pvx, Antoniadis:2020dfq,  Ghilencea:2020piz, Das:2020kff, Gialamas:2020snr, Ghilencea:2020rxc, Iosifidis:2020dck, Bekov:2020dww, Dimopoulos:2020pas,Karam:2021sno, Lykkas:2021vax, Gialamas:2021enw, Antoniadis:2021axu,  Gialamas:2021rpr, AlHallak:2021hwb,  Dioguardi:2021fmr,Dimopoulos:2022tvn, Dimopoulos:2022rdp, Pradisi:2022nmh, Durrer:2022emo, Salvio:2022suk, Antoniadis:2022cqh,  Lahanas:2022mng, Gialamas:2022xtt, Dioguardi:2022oqu,Iosifidis:2022xvp,Gialamas:2023aim, Gialamas:2023flv,SanchezLopez:2023ixx,Dioguardi:2023jwa,DiMarco:2023ncs,Gomes:2023xzk,Hu:2023yjn} is maintained in the presence of non-minimal $f(\phi){\cal{R}}$ couplings~~\cite{Bauer:2008zj, Rasanen:2017ivk, Tenkanen:2017jih, Racioppi:2017spw, Markkanen:2017tun, Jarv:2017azx, Fu:2017iqg, Racioppi:2018zoy, Kozak:2018vlp, Rasanen:2018ihz, Almeida:2018oid, Shimada:2018lnm, Takahashi:2018brt, Jinno:2018jei, Rubio:2019ypq, Racioppi:2019jsp,Shaposhnikov:2020fdv, Borowiec:2020lfx, Jarv:2020qqm, Karam:2020rpa, McDonald:2020lpz, Langvik:2020nrs, Shaposhnikov:2020gts, Shaposhnikov:2020frq, Mikura:2020qhc, Verner:2020gfa, Enckell:2020lvn, Reyimuaji:2020goi, Karam:2021wzz,  Mikura:2021ldx, Racioppi:2021ynx,  Mikura:2021clt, Cheong:2021kyc, Azri:2021uat, Racioppi:2021jai,Piani:2022gon, Karananas:2022byw, Rigouzzo:2022yan, Gialamas:2022gxv,Hyun:2023bkf,Piani:2023aof,Gialamas:2023emn,Rigouzzo:2023sbb,Barman:2023opy}. Nevertheless, the effect of derivative couplings $\sim (\partial\phi)^2{\cal{R}}$  or $\sim (\partial_{ \mu}\phi\partial_{ \nu}\phi)\cR^{ \mu\nu}$ on this behaviour~\cite{Gialamas:2020vto,Nezhad:2023dys}, being an open issue, is the main focus of the present paper. As working examples we consider the case of a scalar field with a quadratic potential and the case of a scalar field with a quartic self interaction. In both models the inclusion of the derivative couplings can increase the value of the spectral index. In the absence of the $\cR^2$ term the tensor-to-scalar ratio, $r$, is reduced but the reduction is not enough to render the models compatible with the observational data. The largest reduction of $r$ comes from the inclusion of the $\cR^2$ term~\cite{Enckell:2018hmo, Antoniadis:2018ywb}. The effect of the derivative couplings in combination with the $\cR^2$ term can rescue both models aligning their inflationary predictions more closely with the latest observational constraints.
We also analyze the issue of reheating in the simpler case of the quadratic model and, without adopting any specific reheating mechanism, we find that the maximum reheating temperature $T_{\rm max}$, is of order $\sim 3 \times 10^{15} \GeV $ , with small deviations for varying the derivative coupling parameter $\tilde{\alpha}$ and small values of the ${\cal{R}}^2$-parameter $\b$. For large $\b$ the maximum reheating temperature is independent of $\tilde{\a}$ and behaves as $\b^{-1/4}$.

The outline of the paper is as follows. In section~\ref{the_model} we set up the theoretical framework of metric-affine gravity with derivative couplings and quadratic in curvature terms. Section~\ref{inflation} discusses the inflationary predictions of the quadratic and quartic models. The reheating temperature is computed in section~\ref{reheating}. Finally, we summarize and conclude in section~\ref{Summary}.

Throughout the paper, we adopt the mostly plus signature for the metric and we use natural units, setting the reduced Planck mass $M_{\rm P}$ to one.

\section{The model}
\label{the_model}
Metric-affine theories of gravity treat the metric tensor $g_{\m\n}$ and the connection $\tns{\Gamma}{^\l_\m_\n}$ as independent variables. The independent connection can be decomposed as
\be
\tns{\Gamma}{^\l_\m_\n} = \{\tns{}{^\l_\m_\n}\} + \tns{\cC}{_\m^\l_\n}\,,
\ee
where $\{\tns{}{^\l_\m_\n}\}$ is the usual Levi-Civita tensor and $\tns{\cC}{^\l_\m_\n}$ is the distortion tensor. The Riemann tensor is given by\footnote{Throughout this paper we employ different symbols to represent the curvature scalar or/and tensors, which in the standard metric gravity we denote by $R$, while in metric-affine gravity by $\mathcal{R} $.}
\begin{align}
\tns{\cR}{^\a_\b_\g_\d} &=\partial_\g \tns{\Gamma}{^\a_\d_\b} - \partial_\d \tns{\Gamma}{^\a_\g_\b} +\tns{\Gamma}{^\a_\g_\m}\tns{\Gamma}{^\m_\d_\b} -\tns{\Gamma}{^\a_\d_\m}\tns{\Gamma}{^\m_\g_\b} \nonumber
\\ 
&= \tns{R}{^\a_\b_\g_\d} +\nabla_\g \tns{\cC}{_\d^\a_\b} - \nabla_\d \tns{\cC}{_\g^\a_\b} +\tns{\cC}{_\g^\a_\m}\tns{\cC}{_\d^\m_\b} -\tns{\cC}{_\d^\a_\m}\tns{\cC}{_\g^\m_\b}\,,
\end{align}
where $\tns{R}{^\a_\b_\g_\d}$ is the metric Riemann tensor constructed from the metric and $\nabla$ is the covariant derivative in terms of the Levi-Civita connection. The only symmetry of $\tns{\cR}{^\a_\b_\g_\d}$  is the antisymmetry under the interchange of the last two indices. As a result, there are three non-zero contractions given by
\be
\tns{\cR}{_\m_\n} = \tns{\cR}{^\r_\m_\r_\n}\,, \qquad \tns{\widehat{\cR}}{^\m_\n} = g^{\a\b} \tns{\cR}{^\m_\a_\b_\n}\,, \qquad \tns{\tilde{\cR}}{_\m_\n} = \tns{\cR}{^\a_\a_\m_\n}\,,
\ee
called the Ricci, co-Ricci, and homothetic curvature tensor, respectively. 
There is a single Ricci scalar determined through an additional contraction of either the Ricci tensor or the co-Ricci tensor, expressed as follows:
\be
\cR = g^{\m\n}\tns{\cR}{_\m_\n} = -\tns{\widehat{\cR}}{^\m_\m}\,.
\ee

 We consider the following metric-affine action of a scalar field $\phi$ coupled non-minimally to the Ricci scalar as well as the Ricci tensors through derivative couplings:
\begin{align}
\label{eq:act_0}
\mathcal{S} = \int {\rm d}^4 x \sqrt{-g} \bigg( \frac12 (f(\phi)+\a_1 X)\cR -\frac12 K(\phi) X +\a_2 \cR^{\m\n}X_{\m\n} +\a_3 \tns{\widehat{\cR}}{^\m^\n}  X_{\m\n} + \frac{\b}{4} \cR^2 -V(\phi) \bigg)\,,
\end{align}
where
\be
X_{\m\n} = \partial_\m \phi \partial_\n \phi \qquad \text{and} \qquad X=g^{\m\n} X_{\m\n}\,.
\ee
There is no coupling of the homothetic curvature Ricci tensor $\tns{\tilde{\cR}}{_\m_\n}$ due to the antisymmetry of its indices. Note that in the metric case (i.e. $\tns{\cC}{_\m^\l_\n} = 0$) there is no separate $a_3$ coupling since $\widehat{R}_{\m\n} = - R_{\m\n}$. The functions $F(\phi)$, $K(\phi)$, and $V(\phi)$ represent nonminimal couplings, non-canonical kinetic terms, and the potential term of the scalar field, respectively. As we have remarked in the introduction quadratic terms of the curvature are bound to be generated from quantum corrections due to gravitating matter fields. Nevertheless, general quadratic terms of the Riemann and Ricci tensors~\cite{Annala:2020cqj,Annala:2021zdt,Gialamas:2021enw} are known to be associated with unphysical degrees of freedom~\cite{BeltranJimenez:2019acz,BeltranJimenez:2020sqf,Marzo:2021iok,Annala:2022gtl,Barker:2024ydb,Barker:2024dhb}, in contrast to quadratic terms of the Ricci scalar which is renowned for its reliable inflationary predictions~\cite{Enckell:2018hmo,Antoniadis:2018ywb} both in the metric as well as the metric-affine case. Therefore, in the action~\eqref{eq:act_0}, we have also incorporated a quadratic Ricci scalar term.  Note that the $\mathcal{R}^2$ term is not the only possible safe quadratic term that can be added to the action. A quadratic term constructed from the Holst invariant  $\tns{\eps}{_\a^\b^\g^\d} \tns{\cR}{^\a_\b_\g_\d}$, can also be incorporated. The implications of such terms in inflation have been investigated in studies such as ~\cite{Salvio:2022suk,Gialamas:2022xtt,DiMarco:2023ncs}. Nevertheless, in the present study we concentrate on the presence of $\cR^2$. 

The $\cR^2$ term can be written in terms of the auxiliary scalar field $\chi$ as $\cR^2 = 2\chi\cR -\chi^2$, so the action~\eqref{eq:act_0} 
takes the form
\begin{align}
\label{eq:act_1}
\mathcal{S} = \int {\rm d}^4 x \sqrt{-g} \bigg( \frac12 (\mathcal{F}(\phi,\chi)+\a_1 X)\cR -\frac12 K(\phi) X +\a_2 \cR^{\m\n}X_{\m\n} +\a_3 \tns{\widehat{\cR}}{^\m^\n}  X_{\m\n}  -U(\phi,\chi) \bigg)\,,
\end{align}
with
\be
\label{eq:r2_rep}
\mathcal{F}(\phi,\chi) = f(\phi) +\b \chi \qquad \text{and} \qquad U(\phi,\chi) = V(\phi) +\frac{\b}{4} \chi^2\,.
\ee

To delve into the dynamics of inflation within the theory, it becomes necessary to rephrase the action in what is known as the Einstein frame. While actions constructed solely from the Ricci scalar can be transformed via a Weyl rescaling (or conformal transformation), since our action ~\eqref{eq:act_0} involves derivative couplings of the scalar field to the Ricci tensors, we need to employ a broader set of transformations, namely the {\textit{disformal transformations}}. These transformations are defined as (we have closely followed the notation of~\cite{Nezhad:2023dys})
\be
g_{\m\n} = \g_1 (\phi,\tilde{X})\tilde{g}_{\m\n} + \g_2(\phi,\tilde{X})X_{\m\n}\,,
\ee
where $\tilde{X} = \tilde{g}^{\m\n} X_{\m\n}$. The inverse transformation is
\be
\tilde{g}_{\m\n} = \tilde{\g}_1 (\phi,X)g_{\m\n} + \tilde{\g}_2(\phi,X)X_{\m\n}\,,
\ee
with $\tilde{\g}_1 =1/\g_1$ and $\tilde{\g}_2 = -\g_2/\g_1$, while the determinants are related by the equation $g = \tilde{g} \g_1^3(\phi,\tilde{X}) ( \g_1(\phi,\tilde{X}) + \g_2(\phi,\tilde{X})\tilde{X})$.
Using the relations for the inverse metrics $g^{\m\n}$ and $\tilde{g}^{\m\n}$ we obtain also that
\be
X=\frac{\tilde{X}}{\g_1(\phi,\tilde{X}) + \g_2(\phi,\tilde{X}) \tilde{X}} \qquad \text{and} \qquad \tilde{X} = \frac{X}{\tilde{\g}_1 (\phi,X) + \tilde{\g}_2 (\phi,X) X}\,. 
\ee
Following~\cite{Nezhad:2023dys} we may replace the co-Ricci tensor with the average Ricci tensor,
defined as $\overline{\cR}_{\m\n} = (\cR_{\m\n}+ \widehat{\cR}_{\m\n})/2$, which vanishes if the connection is metric compatible. In general metric-affine theories have non-zero torsion $T_{\m\l\n} = 2 \cC_{[\m|\l|\n]}$ (see~\cite{Mavromatos:2023wkk} for a recent review on gravitational theories with torsion) and non-metricity $\nabla_\r g_{\m\n} = -2\cC_{(\m\n)\r}$. In the Einstein-Cartan gravity framework, non-metricity is assumed to be zero, whereas in Palatini gravity, the torsion is required to vanish. In the subsequent discussion, we will initially maintain the coupling to the average Ricci tensor. However, ultimately, we will omit it, focusing our analysis solely on the Einstein-Cartan case.

By substituting the average Ricci tensor into equation~\eqref{eq:act_1} and applying the disformal transformation, we derive
% \begin{align}
% \label{eq:act_2}
% \mathcal{S} = \int {\rm d}^4 x &\sqrt{-g} \bigg[\frac{1}{2} (1+\gamma X)^{1/2} \left(\g_1\mathcal{F}(\phi,\chi) +\frac{\a_1 X}{1+\g X} \right)\cR +\a_3 (1+\gamma X)^{-1/2} \overline{\cR}^{\m\n} X_{\m\n} \nonumber
% \\
% & +\frac12 (1+\gamma X)^{-1/2} \left(-\g_2\mathcal{F}(\phi,\chi) +(1+\gamma X)^{-1}(\a_2-\a_3-(\a_1+\a_3)\g X) \right) \cR^{\m\n} X_{\m\n} \nonumber
% \\
% & -\frac12 \g_1  (1+\gamma X)^{-1/2} K(\phi) X -\g_1^2 (1+\g X)^{1/2} U(\phi,\chi) \bigg]\,.
% \end{align}
\begin{align}
\label{eq:act_2}
\mathcal{S} = \int {\rm d}^4 x \sqrt{-g} \bigg[ & F_1(\phi,X,\chi)\frac{\cR}{2} +F_2(\phi,X) \overline{\cR}^{\m\n} X_{\m\n} + F_3(\phi,X,\chi) \cR^{\m\n} X_{\m\n} \nonumber
\\
& -F_4(\phi,X) X -F_5(\phi,X,\chi) U(\phi,\chi) \bigg]\,.
\end{align}
The functions $F_i$ are given by
\begin{subequations}
\begin{align}
F_1(\phi,X,\chi) &= (1+\gamma X)^{1/2} \left(\g_1\mathcal{F}(\phi,\chi) +\frac{\a_1 X}{1+\g X} \right)\,,
\\
F_2(\phi,X) &= \a_3 (1+\gamma X)^{-1/2} \,,
\\
F_3(\phi,X,\chi) &= \frac12 (1+\gamma X)^{-1/2} \left(-\g_2\mathcal{F}(\phi,\chi) +\frac{\a_2-\a_3-(\a_1+\a_3)\g X}{(1+\gamma X)} \right)\,,
\\
F_4(\phi,X) &= \frac12 \g_1  (1+\gamma X)^{-1/2} K(\phi)\,,
\\
F_5(\phi,X,\chi) &= \g_1^2 (1+\g X)^{1/2}\,,
\end{align}
\end{subequations}
where, for brevity, we have omitted the tildes from the rescaled quantities and the arguments from the functions $\g$ and $\g_1$. Our action~\eqref{eq:act_2} aligns with the one presented in~\cite{Nezhad:2023dys}, with the sole distinction being the replacement given by equation~\eqref{eq:r2_rep}. Note however that the presence of the auxiliary scalar $\chi$ will turn out to have important effects on the inflationary behaviour.

If we assume that $\alpha_3=0$ (i.e. the Einstein-Cartan case), to derive the action in the EF, we essentially need to solve the system of equations $F_1(\phi,X,\chi) = 1$ and $F_3(\phi,X,\chi) = 0$. Solving this system is inherently challenging. However, we can approximate the solutions by assuming that in the slow-roll approximation, the higher-order kinetic terms are negligible (i.e. $X\ll 1$), particularly during inflation~\cite{Gialamas:2019nly}, as well as during reheating~\cite{Karam:2021sno}. An approximate solution under this assumption is\footnote{Note that the invertibility of the disformal transformation requires $\g_1>0, \,\,\,\, \g_2 \ge 0, \,\,\,\, \g_1+\tilde{X}\g_2>0, \,\, \newline \tilde{\g}_1 -X \partial \tilde{\g}_1 /\partial X -X^2 \partial \tilde{\g}_2/\partial X\neq 0$. Using the approximate solution~\eqref{eq:appr_sr} we see that the above system is equivalent to $\a_2> 0$,\,\, $\a_1+\a_2<0$ and $|X|< 1/(-\a_1+\a_2/2)$. Additionally, we can infer that the combination $4\alpha_1 + \alpha_2$ involved in the inflationary dynamics is negative, as required. }
\be
\label{eq:appr_sr}
\g \simeq \a_2 -\frac{\a_2^2}{2}X +\frac{5\a_2^3}{8}X^2\,, \quad \g_1\simeq \frac{1}{\mathcal{F}(\phi,\chi)}\left( 1-(\a_1+\a_2/2)X +(\a_1 \a_2 +5\a_2^2/8)X^2\right)\,,
\ee
where we kept terms up to $\mathcal{O}(X^2)$. Substituting the solution back to the action we obtain
\begin{align}
\label{eq:act_3}
\mathcal{S} = \int {\rm d}^4 x \sqrt{-g} \bigg[ & \frac{R}{2} -\frac{K(\phi)X}{2\mathcal{F}(\phi,\chi)} (1-(\a_1+\a_2)X)
\nonumber
\\
&
-\frac{U(\phi,\chi)}{\mathcal{F}^2(\phi,\chi)}\left(1-(2\a_1+\a_2/2)X+(\a_1^2 +2\a_1\a_2 +5 \a_2^2/8)X^2\right) \bigg]\,,
\end{align}
where now the Ricci scalar $R$ is the one constructed by the Levi-Civita connection. Note that for $\a_1=\a_2=0$ the above action is reduced to the known Palatini-$\cR^2$ action~\cite{Enckell:2018hmo,Antoniadis:2018ywb}. 
The subsequent step involves varying the action with respect to the auxiliary field $\chi$. Upon doing so, the solution $\chi = \chi(\phi,X)$ must be re-expanded in powers of the kinetic term $X$ and then substituted back into the action. The resulting effective action will represent the final metric action of the scalar field $\phi$, featuring modified potential and kinetic terms. The variation gives
\be
\frac{\d \mathcal{S}}{\d \chi} = 0 \quad \Rightarrow \quad \chi = \frac{4V(\phi) +A(\phi) X +B(\phi) X^2}{f(\phi) +C(\phi) X +D(\phi) X^2}\,,
\ee
with
\begin{subequations}
\begin{align}
A(\phi) = & K(\phi) f(\phi) -4V(\phi)(2\a_1+\a_2/2)\,, 
\\
B(\phi) = & 4V(\phi) (\a_1^2+2\a_1\a_2+5\a_2^2/8) -(\a_1+\a_2)K(\phi)f(\phi)\,,
\\
C(\phi) = & -\b K(\phi) - f(\phi) (2\a_1+\a_2/2)\,,
\\
D(\phi) = & \b(\a_1+\a_2) K(\phi) +(\a_1^2+2\a_1\a_2+5\a_2^2/8)f(\phi)\,.
\end{align}
\end{subequations}
Expanding in powers of $X$ we obtain\footnote{In the minimal case $f(\phi) = K (\phi) = 1$ the auxiliary field reads
\be
\label{eq:chi_exp_2}
\chi \simeq 4V(\phi) +(1+4\b V(\phi))X +(1+4\b V(\phi))(\b +\a_1-\a_2/2) X^2 +\mathcal{O}(X^3)\,.
\ee}
% \begin{align}
% \label{eq:chi_exp}
% \chi \simeq  &\frac{4V(\phi)}{f(\phi)} +\frac{1}{f(\phi)} \left(A(\phi)-\frac{4V(\phi)C(\phi)}{f(\phi)} \right)X \nonumber
% \\
% &
% -\frac{C(\phi)}{f^2(\phi)} \left(A(\phi) - \frac{4V(\phi)C(\phi)}{f(\phi)}  +\frac{4V(\phi)D(\phi)}{C(\phi)} -\frac{B(\phi) f(\phi)}{C(\phi)} \right) X^2\,.
% \end{align}

\be
\label{eq:chi_exp}
\chi\simeq\frac{4V(\phi)}{f(\phi)}+XK(\phi)\left(1+4\beta\frac{V(\phi)}{f^2(\phi)}\right)
+X^2\frac{K(\phi)}{2f(\phi)}\left(1+4\beta\frac{V(\phi)}{f^2(\phi)}\right)\left(\,(2\alpha_1-\alpha_2)f(\phi)+2\beta K(\phi)\right)\,.
\ee
Substituting the eq.~\eqref{eq:chi_exp} back to the action~\eqref{eq:act_3} and re-expanded in powers of $X$ the $\chi$-dependent part of the first line gives
\be
\label{eq:Lkin}
{\cal{L}}_1\simeq-X\frac{K(\phi)}{2f(\phi)\left(1+4\beta\frac{V(\phi)}{f^2(\phi)}\right)}+X^2\frac{K(\phi)}{2f^2(\phi)\left(1+4\beta\frac{V(\phi)}{f^2(\phi)}\right)}\left((\alpha_1+\alpha_2)f(\phi)+\beta K(\phi)\right)\,,
\ee
% {\color{red}{\be\mathcal{L}_1 = -\frac{X}{2}\left( \frac{f(\phi)}{f^2(\phi)+4\b V(\phi)}\right) + \frac{X^2}{4} \left(\frac{2(\a_1+\a_2)f(\phi)+2\b}{f^2(\phi)+4\b V(\phi)}\right) \ee}}
while the second line reads
\be 
{\cal{L}}_2\simeq\frac{V(\phi)}{f^2(\phi)+4\beta V(\phi)}-\frac{X}{2}\frac{(4\alpha_1+\alpha_2)V(\phi)}{f^2(\phi)+4\beta V(\phi)}
+\frac{X^2}{8}\frac{\left(2\beta K^2(\phi)+(8\alpha_1^2+5\alpha_2^2+16\alpha_1\alpha_2)V(\phi)\right)}{f^2(\phi)+4\beta V(\phi)}\,.
\ee
Having done all the groundwork, we can substitute the last expansions in the action~\eqref{eq:act_3} to obtain
\be
\mathcal{S}=\int {\rm d}^4x \sqrt{-g} \left(\frac{R}{2} -\bar{K}(\phi)\frac{X}{2} -\bar{U}(\phi) + \mathcal{O}(X^2) \right)\,,
\ee
with 
\be 
\Bar{K}(\phi)=\frac{ -\Tilde{\a} V(\phi)+f(\phi) K(\phi)}{ \left(f^2(\phi)+4 \beta  V(\phi)\right)} \qquad \text{and} \qquad \Bar{U}(\phi)=\frac{V(\phi)}{f^2(\phi)+4\beta V(\phi)}\,,
\ee
where we have defined $4\a_1+\a_2 = \tilde{\a}$. Although we have retained the functions $f(\phi)$ and $K(\phi)$ for generality in this discussion, moving forward, we will simplify the analysis by considering the minimal case where $f(\phi) = K(\phi) = 1$. Note that, achieving canonical form for the kinetic term is possible through the field redefinition ${\rm d}\phi_{c} = \sqrt{\bar{K}(\phi)} {\rm d}\phi $. Consequently, the canonically normalized inflaton $\phi_{c}$ can, in principle, be expressed as a function of $\phi$. However, it is not obligatory to employ a canonical field for determining inflationary parameters. By directly working with $\phi$, this impediment can be bypassed.

\section{Inflation}
\label{inflation}
Regarding cosmological observables and assuming the slow-roll approximation, we initiate the discussion by introducing the scalar ($\mathcal{P}_\zeta $) and tensor ($\mathcal{P}_T$) power spectra, crucial elements in inflationary cosmology. By selecting an arbitrary pivot scale $k_\star$ that exited the horizon, the expressions for the scalar and tensor power spectra take the form:
\be
\label{eq:spectra}
\mathcal{P}_\zeta (k)=A_s \left(\frac{k}{k_\star} \right)^{n_s -1}, \quad  A_s\simeq\frac{1}{24\pi^2}\frac{\bar{U}(\phi_\star)}{\epsilon_{\bar{U}}(\phi_\star)}  \qquad \text{and} \qquad \mathcal{P}_T (k)\simeq\frac{2\bar{U}(\phi_\star)}{3\pi^2} \left(\frac{k}{k_\star} \right)^{n_t}\,,
\ee
where $A_s$ is the amplitude of the power spectrum of scalar perturbations.
The scalar ($n_s$) and tensor ($n_t$) spectral indices given by
\be
\label{eq:index}
n_s-1=\frac{{\rm d} \ln \mathcal{P}_\zeta (k) }{{\rm d} \ln k} \simeq -6\epsilon_{\bar{U}} +2\eta_{\bar{U}}  \qquad \text{and} \qquad n_t= \frac{{\rm d} \ln \mathcal{P}_T (k) }{{\rm d} \ln k}\,,
\ee
characterize the scale-dependence of the power spectra~\eqref{eq:spectra}. In the equations above we have used the potential slow-roll parameters 
\be
\label{eq:pslp}
\eps_{\bar{U}} = \frac{1}{2\bar{K}(\phi)} \left( \frac{\bar{U}'(\phi)}{\bar{U}(\phi)} \right)^2 \qquad \text{and} \qquad \eta_{\bar{U}} = \frac{\left(\bar{K}^{-1/2}(\phi) \bar{U}'(\phi)\right)'}{\bar{K}^{1/2}(\phi)\bar{U}(\phi)}\,.
\ee
In these equations primes denote derivatives with respect the scalar field, while
both slow-roll parameters are small $(\ll 1)$ during inflation and one of them approaches unity near its end. 

The tensor-to-scalar ratio is defined as
\be
\label{eq:ttsr}
    r= \frac{\mathcal{P}_T (k)}{\mathcal{P}_\zeta (k)} \simeq 16\epsilon_{\bar{U}},
\ee
while the duration of inflation is measured by the number of $e$-folds
\be
\label{eq:efolds}
N_\star =\int^{\phi_{\star}}_{\phi_{\rm end}} \bar{K}(\phi) \frac{\bar{U}(\phi)}{\bar{U}'(\phi)}{\rm d}\phi\,.
\ee

The inflationary predictions are significantly constrained by observations of the cosmic microwave background (CMB), as demonstrated in~\cite{Planck:2018jri,BICEP:2021xfz,Tristram:2021tvh,Galloni:2022mok}. The most recent combination of Planck, BICEP/Keck, and BAO data has established the following limits on the observable values\footnote{To refine our analysis, the constraints obtained from the Planck, BICEP/Keck, and BAO data yield $r<0.036$ in ~\cite{BICEP:2021xfz}, $r<0.032$ in~\cite{Tristram:2021tvh}, and $r<0.028$ in~\cite{Galloni:2022mok}. In this paper, we adopt the approximate value $r<0.03$.} at the pivot scale $k_\star = 0.05\, {\rm Mpc}^{-1}$:
\be\label{eq:constraints}
    A_s = (2.10\pm 0.03)\times10^{-9},\qquad n_s=0.9649 \pm 0.0042  \quad (1\sigma\mbox{ region}), \qquad r < 0.03\,.
\ee
Subsequently, we examine particular models and delve into their predictions. We focus on an intriguing category of models where the potential, $V$, takes the form of a monomial in the field $\phi$, i.e. $V\sim \phi^n$, with $n$ even integer. More precisely we will study the quadratic $(n=2)$ and quartic $(n=4)$ models of inflation.

\subsection{The quadratic model}
\label{sec:inf_quad}
We first consider the simple case of the minimally coupled ($f(\phi)=1$) quadratic model with a potential
\be
V(\phi) = \frac{m^2}{2}\phi^2\,,
\ee
where $m^2$ is a parameter of dimension ${\rm mass}^2$.
Provided that the $\mathcal{O}(X^2)$ terns are small, the first slow-roll parameters~\eqref{eq:pslp} are given by
\be
\eps_{\bar{U}} = \frac{4}{\phi^2(2-\tilde{\a}m^2\phi^2)(1+2\b m^2 \phi^2)}\,, \qquad \eta_{\bar{U}} =  \frac{8\left(1 +\b m^2\phi^2(3\tilde{\a}m^2\phi^2 -4)\right)}{\phi^2(2-\tilde{\a}m^2\phi^2)^2(1+2\b m^2 \phi^2)}\,.
\ee
The number of $e-$folds~\eqref{eq:efolds}, left to the end of on inflation are
\be
\label{eq:Nstar}
N_\star = \frac{1}{16}(\phi_{\rm end}^2 - \phi_{\star}^2) \left(\tilde{\a}m^2(\phi_{\rm end}^2 + \phi_{\star}^2) -4\right)  \simeq \frac{1}{16} \phi_{\star}^2 (4-\tilde{\a}m^2 \phi_{\star}^2)\,,
\ee
where the second equality holds for $\phi_{\rm end}^2 \ll \phi_{\star}^2$.
The above equation (without neglecting $\phi_{\rm end}$ is a quadratic equation for $\phi_{\star}^2$. The sole solution that restores the correct $\tilde{\a}=0$ limit is
\be
\label{eq:phistar}
\phi_\star^2 =  \frac{2-\sqrt{\tilde{\a}^2m^4\phi_{\rm end}^4-4\tilde{\a}m^2\phi_{\rm end}^2+4-16\tilde{\a}m^2 N_\star}}{\tilde{\a}m^2} \xrightarrow{\tilde{a}\rightarrow 0} \phi_{\rm end}^2+4N_\star\,.
\ee

The field value at the end of inflation is defined by the condition $\eps_{\bar{U}} (\phi_{\rm end})= 1 \Rightarrow$
\be
\label{eq:phi_end}
2\tilde{\a}\b m^4\phi_{\rm end}^6 +(\tilde{\a}m^2 -4\b m^2)\phi_{\rm end}^4 -2\phi_{\rm end}^2 + 4 =0\,.
\ee
In~\cite{Gialamas:2019nly}, it has been shown that for $\tilde{\a} = 0$ the field value at the end of inflation is bounded from above, namely $\phi_{\rm end}^2 < 2$. So in this case, the condition $\phi_{\rm end}^2 \ll \phi_{\star}^2$ is indeed true. In our scenario we have seen numerically that such a bound for the $\phi_{\rm end}^2$ also exists, thereby ensuring that the approximation $\phi_{\rm end}^2 \ll \phi_{\star}^2$ holds true.

%%%%%%%%%%%%%%%%%%%%%%%%%%FIGURE%%%%%%%%%%%%%%%%%%%%%%%%%%%%%%%%
\begin{figure}[t]
\centering
\includegraphics[width=0.49\textwidth]{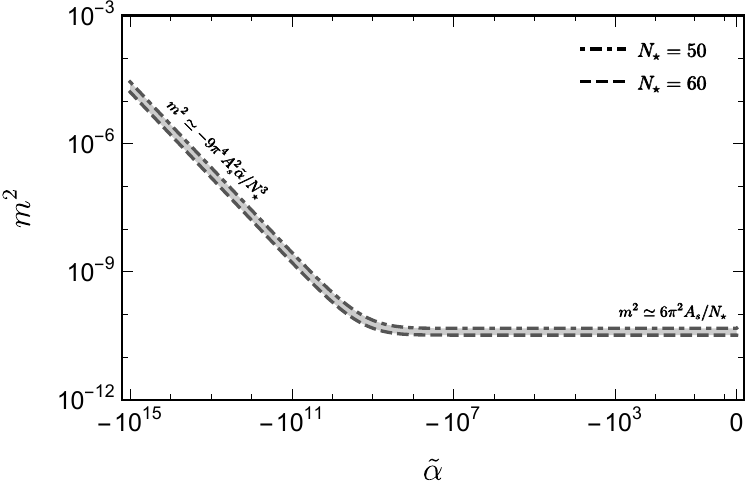}
\includegraphics[width=0.49\textwidth]{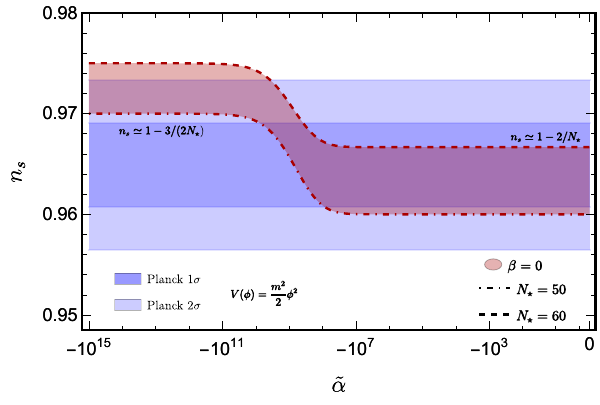}
\caption{{\textbf{Left:}} The mass parameter $m^2$ given by eq.~\eqref{eq:m2} as function of the parameter $\tilde{\a}$. {\textbf{Right:}} The spectral index, given by eq.~\eqref{eq:ns}, as function of the parameter $\tilde{\a}$ for $\b=0$. The shaded regions represent the permissible parameter space at confidence levels of  $68\%$ (dark blue) and $95\%$ (light blue), as derived from the most recent combination of Planck, BICEP/Keck, and BAO data~\cite{Planck:2018jri,BICEP:2021xfz,Tristram:2021tvh,Galloni:2022mok}.} 
\label{fig:1}
\end{figure}
%%%%%%%%%%%%%%%%%%%%%%%%%FIGURE_END%%%%%%%%%%%%%%%%%%%%%%%%%%%%%%%%%
%%%%%%%%%%%%%%%%%%%%%%%%%%FIGURE%%%%%%%%%%%%%%%%%%%%%%%%%%%%%%%%
In what follows, we safely omit the term $\phi_{\rm end}$. Under this approximation the field value at the horizon crossing is given by
\be
\label{eq:phistar_ap}
\phi_\star^2 \simeq   \frac{2-2\sqrt{1-4\tilde{\a}m^2 N_\star}}{\tilde{\a}m^2}\,.
\ee
Using $\phi_\star$, given above, $A_s$ of eq.~\eqref{eq:spectra} is written in terms of $N_\star$ as
\be
\label{eq:As_apr}
A_s \simeq \frac{\sqrt{1-4\tilde{\a}m^2N_\star}\left(1-\sqrt{1-4\tilde{\a}m^2N_\star}\right)^2}{24\pi^2\tilde{\a}^2m^2}\,,
\ee
which under this approximation does not depend on the parameter $\b$.
The above equation can be used in order to fix one of the parameters. If we solve with respect to $m^2$, the only solution that recovers the correct $\tilde{\a}=0$ limit, namely $m^2\simeq 6\pi^2 A_s/N_\star^2$, is
\be
\label{eq:m2}
\hspace{-0.3cm}m^2 = \frac{3\pi^2A_s}{4\mathring{A}N_\star^3}\left[N_\star^2+4\mathring{A} N_\star -\mathring{A}^2 -(N_\star -\mathring{A}) \left(N_\star^2 -6\mathring{A} N_\star +\mathring{A}^2 \right)^{1/2}  \right]\,,\quad \mathring{A}=6\pi^2A_s\tilde{\a}\,.
\ee
The impact of the parameter $\tilde{\a}$ on $m^2$ is visible once the parameter $\mathring{A}$ reaches a comparable or larger magnitude than $N_\star$.\footnote{This can be easily observed if we expand around $\mathring{A}\sim 0$. The mass for small $|\mathring{A}|$ is given by 
\be
m^2 \simeq \frac{6\pi^2 A_s}{N_\star^2}\left[ 1+ \left(\frac{\mathring{A}}{N_\star}\right)^2+ 4\left(\frac{\mathring{A}}{N_\star}\right)^3 +\mathcal{O}(\mathring{A}^4/N_\star^4)\right]\,. 
\ee} For $N_\star \sim 50-60$ this occurs for $|\tilde{\a}|\gtrsim 10^8$. 

The left panel of Fig.~\ref{fig:1} illustrates the dependence of the parameter $m^2$ on the parameter $\tilde{\a}$ according to the aforementioned equation. It is evident that for ``small" values of $|\tilde{\a}|$ there is no discernible effect on the value of the parameter $m^2$. However, for $|\tilde{\a}|\gg 10^8$ it exhibits a linear growth as
$m^2\simeq -\frac{9\pi^4A_s^2\tilde{\alpha}}{N_{\star}^3}\left[1+\mathcal{O}(N_{\star}/\mathring{A})\right]$. Given that the parameter $m$ represents the mass of the scalar field $\phi$ it is essential for it to remain sub-Planckian. Following this requirement, we derive an upper limit for the parameter $|\tilde{\alpha}|$ given by
\be
\label{eq:bounda1}
|\tilde{\a}| \lesssim 4.3\times 10^{19} \left(\frac{N_\star}{55}\right)^3\,.
\ee

The spectral index~\eqref{eq:index} is given by
\begin{equation}
\label{eq:ns}
n_s \simeq 1 - \frac{1+\sqrt{1-4\tilde{\a}m^2N_\star}-6\tilde{\a}m^2N_\star}{N_\star (1-4\tilde{\a}m^2N_\star)} \simeq
    \begin{cases}
   \displaystyle     1 -\frac{2}{N_\star}\,, & \text{if }\,\, |\mathring{A}|/N_\star\ll 1\\[0.3cm]
   \displaystyle     1 -\frac{3}{2N_\star}\,, & \text{if }\, \, |\mathring{A}|/N_\star\gg 1\,.
    \end{cases}
\end{equation}
As in the $\tilde{\a}=0$ case~\cite{Enckell:2018hmo}, the spectral index is $\b$-independent to leading
order in the slow-roll parameters as well as the scalar power spectrum and the number of $e$-folds. Therefore, for small $|\mathring{A}|$, always compared to $N_\star$, the prediction aligns with that of the simple quadratic model of inflation. As $|\mathring{A}|$ increases the spectral index also increases  eventually reaching the asymptotic value $n_s\simeq 1-3/(2N_\star)$. As is also depicted in the right panel of Fig.~\ref{fig:1} the asymptotic region for large $|\mathring{A}|$ is marginally outside of the $2\s$ observational bounds, namely $N_\star$ is forced to be $\lesssim56$.

\begin{figure}[t!]
\centering
\includegraphics[width=0.48\textwidth]{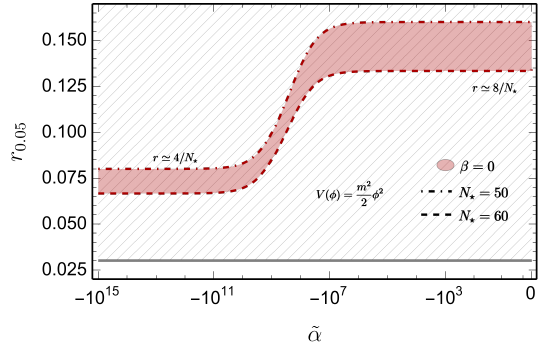}\hspace{0.1cm}
\includegraphics[width=0.49\textwidth]{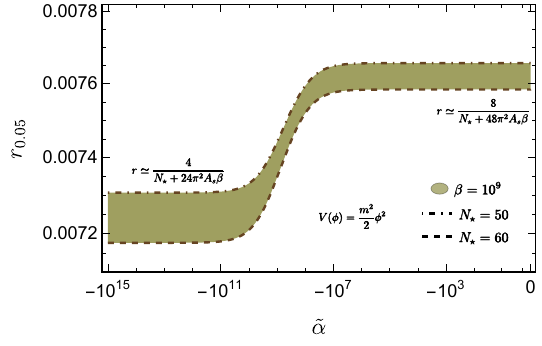}
\caption{The tensor-to-scalar ratio, given by eq~\eqref{eq:r}, as function of the parameter $\tilde{\a}$ for $\b=0$ (left) and $\b=10^9$ (right). The latest observational data excludes the shaded region in the left panel, since $r_{0.05} < 0.03$ at $95\%$ confidence~\cite{BICEP:2021xfz,Tristram:2021tvh,Galloni:2022mok}.} 
\label{fig:2}
\end{figure}

Finally, the tensor-to-scalar ratio~\eqref{eq:ttsr} is given by
\be
\label{eq:r}
r \simeq \frac{16\tilde{\a}m^2}{\sqrt{1-4\tilde{\a}m^2N_\star}(\sqrt{1-4\tilde{\a}m^2N_\star}-1) \left[4\b(\sqrt{1-4\tilde{\a}m^2N_\star}-1)-\tilde{\a}  \right]}\,,
\ee
while its limiting cases are
\be
r\simeq \begin{cases}
   \displaystyle     \frac{8}{N_\star+48\pi^2 A_s\b}\,, & \text{if }\,\, |\mathring{A}|/N_\star\ll 1\\[0.3cm]
   \displaystyle     \frac{4}{N_\star+24\pi^2 A_s\b}\,, & \text{if }\, \, |\mathring{A}|/N_\star\gg 1\,.
    \end{cases}
\ee
As shown in Fig.~\ref{fig:2}, the decrease in the tensor-to-scalar ratio for large $|\mathring{A}|$ alone is insufficient to bring the value of $r$ within the observational limit of $ r<0.03$. As highlighted in the right panel, the introduction of a substantial $\beta$ parameter ($\beta \gtrsim 10^8$) becomes necessary. 
%%%%%%%%%%%%%%%%%%%%%%%%%%FIGURE%%%%%%%%%%%%%%%%%%%%%%%%%%%%%%%%
%%%%%%%%%%%%%%%%%%%%%%%%%FIGURE_END%%%%%%%%%%%%%%%%%%%%%%%%%%%%%%%%%
%%%%%%%%%%%%%%%%%%%%%%%%%%FIGURE%%%%%%%%%%%%%%%%%%%%%%%%%%%%%%%%

\subsection{The quartic model}
As a second model worth studying, is the quartic model
\be
V(\phi) = \frac{\l}{4}\phi^4\,.
\ee
Here, the dimensionless coupling parameter $\l$ is referred to as the quartic coupling.
However, the inflationary predictions of this model do not align with observations, whether considering the pure $\phi^4$ model or with the inclusion of the $\mathcal{R}^2$ term. In the former case, both the tensor-to-scalar ratio and the spectral index deviate from the established bounds. In the latter case, the inclusion of the $\mathcal{R}^2$ term succeeds in reducing the value of $r$, while $n_s$ remains around $\sim 0.94-0.96$  for $N_\star= 50-60$. Subsequently, we will explore how derivative couplings may enhance the value of $n_s$ to bring it within the observationally allowed range\footnote{It is worth noting that the quartic model can be rescued if one assumes a non-minimal coupling of the form $\sim\xi \phi^2 \mathcal{R}$ both in the metric~\cite{Bezrukov:2007ep} and the Palatini~\cite{Bauer:2008zj} formulation. Nevertheless, here we shall consider the minimally coupled case of $f(\phi)=1$ to isolate the effect of a non-minimal derivative coupling on predictions.}.

As in the quadratic model, the first slow-roll parameters, are easily computed to be
\be
\eps_{\bar{U}} = \frac{32}{\phi^2(4-\tilde{\a}\l\phi^4)(1+\b \l \phi^4)}\,, \qquad \eta_{\bar{U}} =  \frac{16\left[12(1-\b\l\phi^4)+\tilde{\a}\l\phi^4(5\b\l\phi^4-1)\right]}{\phi^2(4-\tilde{\a}\l\phi^4)^2(1+\b \l \phi^4)}\,,
\ee
while the number of $e-$folds left to the end of inflation are
\be
N_\star =\frac{1}{96}\left[12(\phi_\star^2-\phi_{\rm end}^2)+\tilde{\a}\l (\phi_{\rm end}^6-\phi_\star^6) \right]\,.
\ee
Using again the approximation $\phi_{\rm end}^2 \ll \phi_{\star}^2$ we obtain that the field value at the horizon crossing is
\be
\phi_\star^2 \simeq \frac{-2(\tilde{\a}\l+Y^2)}{\tilde{\a}\l Y}\,,\qquad \text{with} \qquad Y =\left(6\tilde{\a}^2\l^2 N_\star^2  +\sqrt{\tilde{\a}^3\l^3 (36\tilde{\a}\l N_\star^2-1)} \right)^{1/3}\,.
\ee
%%%%%%%%%%%%%%%%%%%%%%%%%%FIGURE%%%%%%%%%%%%%%%%%%%%%%%%%%%%%%%%
\begin{figure}[t]
\centering
\includegraphics[width=0.49\textwidth]{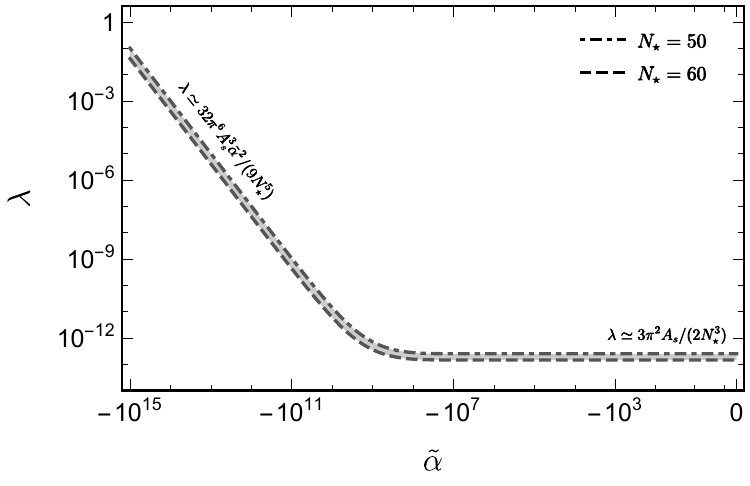}
\includegraphics[width=0.49\textwidth]{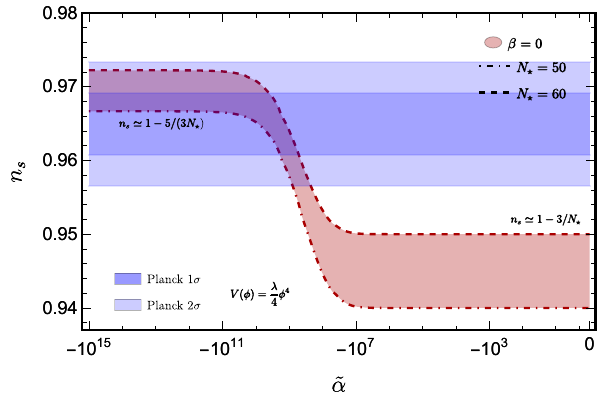}
\caption{{\textbf{Left:}} The quartic coupling $\l$ as function of the parameter $\tilde{\a}$. {\textbf{Right:}} The spectral index, given by eq.~\eqref{eq:ns_4}, as function of the parameter $\tilde{\a}$ for $\b=0$. The shaded regions represent the permissible parameter space at confidence levels of  $68\%$ (dark blue) and $95\%$ (light blue), as derived from the most recent combination of Planck, BICEP/Keck, and BAO data~\cite{Planck:2018jri,BICEP:2021xfz,Tristram:2021tvh,Galloni:2022mok}.} 
\label{fig:3}
\end{figure}
%%%%%%%%%%%%%%%%%%%%%%%%%FIGURE_END%%%%%%%%%%%%%%%%%%%%%%%%%%%%%%%%%
%%%%%%%%%%%%%%%%%%%%%%%%%%FIGURE%%%%%%%%%%%%%%%%%%%%%%%%%%%%%%%%
In contrast to the quadratic model, the intricate expression of $\phi_\star^2$ prevents us from presenting a concise formula for the quartic coupling $\lambda$, akin to what was done for the mass parameter in equation~\eqref{eq:m2}. However, we can provide approximate limiting expressions using the fact that the value of the quartic coupling $\l$ is fixed by the observed value of the amplitude of the scalar power spectrum, $A_s \simeq 2.1\times 10^{-9}$ at the pivot scale $k_\star = 0.05\, {\rm Mpc}^{-1}$. As depicted in the left panel of Fig.~\ref{fig:3}, for $|\tilde{\alpha}|\ll 10^8$, we recover the familiar expression for the quartic model, specifically $\lambda \approx 3\pi^2 A_s/(2N_\star^3)$. On the other hand, for $|\tilde{\alpha}|\gg 10^8$, the quartic coupling increases as $\lambda \approx 32\pi^6 A_s^3 |\tilde{\alpha}|^2/(9N_\star^5)$.  Similar to our approach for the quadratic model and the derivation of the bound~\eqref{eq:bounda1}, imposing a perturbativity condition on the self-quartic coupling $\lambda$ leads to a more stringent upper bound given by
\be
\label{eq:bounda2}
|\tilde{\a}| \lesssim 4.0\times 10^{15} \left(\frac{N_\star}{55}\right)^{5/2}\,.
\ee
It is noteworthy that if we identify the quartic coupling $\lambda$ with the Higgs quartic coupling at the electroweak scale, $\lambda \simeq 0.13$, we find that $|\tilde{\alpha}| \simeq 1.4 \times 10^{15}$ for $N_{\star} = 55$.

Furthermore, since an analytic expression for the quartic coupling $\lambda$ is unavailable, the spectral index is expressed as a function of it, given by:
\begin{align}
\label{eq:ns_4}
n_s\simeq &- \tilde{\a}\l\big[ -\tilde{\a}\l Y^2 +\tilde{\a}^2\l^2 (1+N_\star(48N_\star Y^2+2Y-1)) +12\tilde{\a}^3\l^3 N_\star^2 (3N_\star-5) \nonumber
\\ 
&+(1-8N_\star Y)Y (Y^3-6\tilde{\a}^2\l^2 N_\star^2)\big]/ (Y^3-6\tilde{\a}^2\l^2 N_\star^2)^{2}\,.
\end{align}
The limiting cases are given by
\be
n_s\simeq \begin{cases}
   \displaystyle     1-\frac{3}{N_\star}\,, & \text{if }\,\, |\tilde{\a}|\ll 10^8\\[0.3cm]
   \displaystyle     1-\frac{5}{3N_\star}\,, & \text{if }\, \, |\tilde{\a}|\gg 10^8\,.
    \end{cases}
\ee
The $|\tilde{\alpha}|\ll 10^8$ limit, lies outside of the observational bounds, since it coincides with the pure $\phi^4$ model. In the $|\tilde{\alpha}|\gg 10^8$ limit, the spectral index takes the form $n_s \simeq 1 - 5/(3N_\star) \simeq 0.967-0.972$ for $N_\star =50-60$ (see the right panel of Fig.~\ref{fig:3}). These values fall well within observational bounds, marking a significant achievement as the derivative couplings, in conjunction with $\mathcal{R}^2$, rescue the quartic model of inflation.
%%%%%%%%%%%%%%%%%%%%%%%%%%FIGURE%%%%%%%%%%%%%%%%%%%%%%%%%%%%%%%%
\begin{figure}[t!]
\centering
\includegraphics[width=0.48\textwidth]{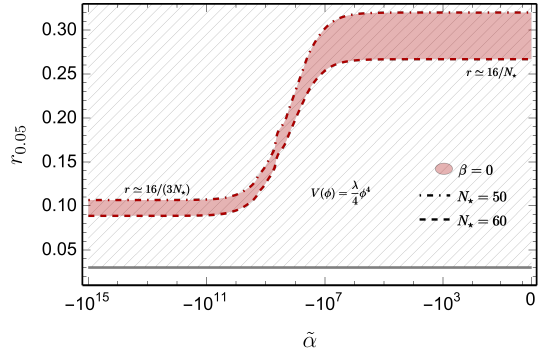}\hspace{0.1cm}
\includegraphics[width=0.49\textwidth]{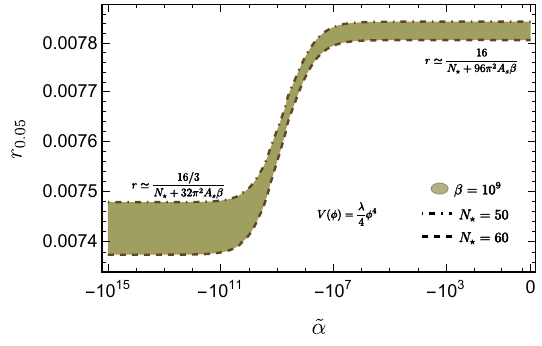}
\caption{The tensor-to-scalar ratio, given by eq~\eqref{eq:r_4}, as function of the parameter $\tilde{\a}$ for $\b=0$ (left) and $\b=10^9$ (right). The latest observational data excludes the shaded region in the left panel, since $r_{0.05} < 0.03$ at $95\%$ confidence~\cite{BICEP:2021xfz,Tristram:2021tvh,Galloni:2022mok}}. 
\label{fig:4}
\end{figure}
%%%%%%%%%%%%%%%%%%%%%%%%%FIGURE_END%%%%%%%%%%%%%%%%%%%%%%%%%%%%%%%%%
%%%%%%%%%%%%%%%%%%%%%%%%%%FIGURE%%%%%%%%%%%%%%%%%%%%%%%%%%%%%%%%
As previously indicated, the derivative couplings alone are insufficient to rescue this model. The tensor-to-scalar ratio is given by
\be
\label{eq:r_4}
r \simeq 64\tilde{\a}^4\l^3Y^5\left[(Y^2+\tilde{\a}\l)(Y^4+\tilde{\a}\l Y^2+\tilde{\a}^2\l^2)(4\b Y^4 +\tilde{\a}\l (\tilde{\a}+8\b)Y^2 + 4\tilde{\a}^2\l^2\b) \right]^{-1}\,,
\ee
with the limits being
\be
r\simeq \begin{cases}
   \displaystyle     \frac{16}{N_\star+96\pi^2 A_s\b}\,, & \text{if }\,\, |\tilde{\a}|\ll 10^8\\[0.3cm]
   \displaystyle     \frac{16/3}{N_\star+32\pi^2 A_s\b}\,, & \text{if }\, \, |\tilde{\a}|\gg 10^8\,.
    \end{cases}
\ee
As depicted in the left panel of Fig.~\ref{fig:4}, the tensor-to-scalar ratio is reduced by approximately $ 67\%$, which is insufficient for agreement with the observational bound $r < 0.03$. The inclusion of $\mathcal{R}^2$ is inevitable, and a value of $\beta >10^8$ (see the right panel of Fig.~\ref{fig:4}) is adequate to relocate the tensor-to-scalar ratio within the allowed region.

%%%%%%%%%%%%%%%%%%%%%%%%%%%%%%%%%%%%%%%%%%%%%%%%%%%%%%%%
\section{Reheating}
\label{reheating}
%%%%%%%%%%%%%%%%%%%%%%%%%%%%%%%%%%%%%%%%%%%%%%%%%%%%%%%%
As an illustrative example within this section, we will compute the reheating temperature, for the quadratic model discussed in Section~\ref{sec:inf_quad}, giving analytic approximations in regions of the parameter space.

In the following, although we do not propose a specific reheating mechanism associated with the decay of the scalar field, we will calculate the maximum\footnote{Regarding Palatini inflationary models, the authors of~\cite{Gialamas:2019nly,Das:2020kff,Lykkas:2021vax,Cheong:2021kyc,Lahanas:2022mng,Zhang:2023hfx} have investigated the inflationary predictions a range of inflationary models. The resulting reheating temperature, exhibits significant variability, ranging from relatively low values near those during BBN, $\sim \mathrm{MeV}$, up to much higher values of around $\sim 10^{16}\,\mathrm{GeV}$.} reheating temperature given by
\be
\label{eq:T_reh}
 T_{\rm max} \equiv \left( \frac{30}{\pi^2} \frac{\rho_{\rm end}}{g_\star(T_{\rm max})}\right)^{1/4}\,,
\ee
where the subscript ``end" in the energy density $\rho$ indicates that it is evaluated at the end of inflation. It is worth noting that $g_\star(T_{\text{max}})$ represents the effective degrees of freedom of entropy density, which is approximately $106.75$ under the assumption of Standard Model particle content and temperatures around $\sim 1 \TeV$ or higher. This value is employed in our calculations.

To estimate the maximum reheating temperature, we only require the value of the energy density at the end of inflation, which can be approximated as\footnote{This equality holds precisely when there are no higher-order kinetic terms. References~\cite{Gialamas:2019nly,Lahanas:2022mng} have considered the inclusion of higher-order kinetic terms, revealing that they yield only a negligible correction, as demonstrated therein.} $\r_{\rm end} \simeq \frac{3}{2}\bar{U}(\phi_{\rm end})\,.$ The field value at the end of inflation is determined by solving equation~\eqref{eq:phi_end}. Since, this is a cubic equation for $\phi_{\rm end}^2$ its solution and consequently the $T_{\rm max}$ are too complicated to be presented, although analytic expression for the unique positive solution does exist. Therefore, our objective is to provide analytic expressions for the limits $|\mathring{A}|/N_\star\ll 1$ and $|\mathring{A}|/N_\star\gg 1$, as described in the preceding section, as well as for the cases $\b \ll 10^9$ and $\b \gg 10^9$.
%%%%%%%%%%%%%%%%%%%%%%%%%%FIGURE%%%%%%%%%%%%%%%%%%%%%%%%%%%%%%%%
\begin{figure}[t!]
\centering
\includegraphics[width=0.50\textwidth]{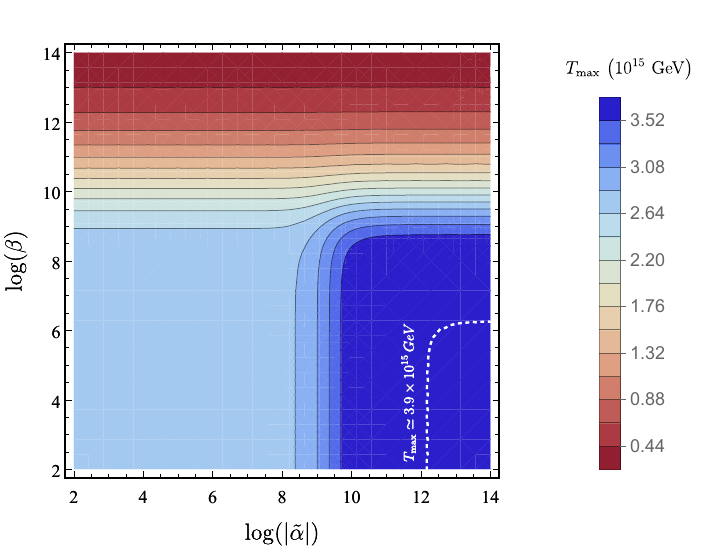}
\caption{In the $\b$, $|\tilde{\a}|$ plane we display the maximum reheating temperature, as given by eq.~\eqref{eq:T_reh}, for $N_\star = 55$.} 
\label{fig:5}
\end{figure}
%%%%%%%%%%%%%%%%%%%%%%%%%FIGURE_END%%%%%%%%%%%%%%%%%%%%%%%%%%%%%%%%%
%%%%%%%%%%%%%%%%%%%%%%%%%%FIGURE%%%%%%%%%%%%%%%%%%%%%%%%%%%%%%%%

For small values of $\b$, i.e. $\b \ll 10^9$ the maximun reheating temperature can be approximated as
\be
\label{eq:Tins_sm_beta}
T_{\rm max} \simeq \begin{cases}
   \displaystyle    0.38\times\left( \frac{1-\sqrt{1-4\mathring{A}/N_\star^2}}{\tilde{\a}} \right)^{1/4} \stackrel{|\tilde{\a}|\rightarrow 0}{\simeq} 2.8\times 10^{15} \left(\frac{55}{N_\star}\right)^{1/2} \GeV\,, & \text{if }\,\, |\mathring{A}|/N_\star\ll 1\\[0.3cm]
   \displaystyle     0.38\times\left( \frac{1-\sqrt{1+\mathring{A}^2/N_\star^3}}{\tilde{\a}} \right)^{1/4} \stackrel{|\tilde{\a}|\rightarrow \infty}{\simeq} 3.9\times 10^{15} \left(\frac{55}{N_\star}\right)^{3/8} \GeV\,, & \text{if }\, \, |\mathring{A}|/N_\star\gg 1\,.
    \end{cases}
\ee
For large values of $\b$, i.e. $\b \gg 10^9$ in both $\tilde{\a} $ regimes we have that
\be
\label{eq:Tins_la_beta}
T_{\rm max} \simeq 7.8\times 10^{17} \beta^{-1/4} \GeV\,.
\ee
Note that we have reinstated the units by multiplying with $M_{\rm Pl} \simeq 2.4\times 10^{18} \GeV$.

In Fig.~\ref{fig:5} we display the maximum reheating temperature, as given by eq.~\eqref{eq:T_reh}, for $N_\star = 55$. The light and dark blue regions represent the limits outlined by equation~\eqref{eq:Tins_sm_beta}, with the white-dotted contour indicating the highest temperature $\sim 3.9 \times 10^{15} \GeV$. As $\beta$ increases, the approximation~\eqref{eq:Tins_sm_beta} becomes more accurate, with the temperature decreasing as $\beta^{-1/4}$ in agreement with~\cite{Gialamas:2019nly,Lahanas:2022mng}.
 In conclusion, as indicated by the figure and the approximate expressions, the maximum reheating temperature is attained for small values of the parameter $\beta$, typically ranging from approximately $(2.8-3.9) \times 10^{15} \GeV$ (for $N_\star = 55$), as $\tilde{\a}$ increases. Conversely, for large values of the parameter $\beta$, it diminishes as $\beta^{-1/4}$, regardless of the value of $\tilde{\alpha}$.

\section{Summary}
\label{Summary}
In the present paper we considered theories of a scalar field, coupled to gravity through non-minimal couplings to curvature that include derivatives of the field, in the general framework of metric-affine theories incorporating quadratic Ricci scalar curvature terms. We employed disformal transformations of the metric to transform the action into the Einstein frame, focusing on the Einstein-Cartan case that allows derivative couplings to the Ricci-tensor and the Ricci scalar. We proceeded to study the inflationary predictions in the simple cases of two models, namely that of a quadratic potential and that of a quartic self-interacting scalar. We find that in both models the effect of the derivative couplings in combination with the quadratic Ricci curvature term leads to predictions aligned with the latest observational data. Derivative couplings tend in general to increase the spectral index, while a reduction to the tensor-to-scalar ration comes mostly from the quadratic Ricci scalar term.  Natural considerations regarding the perturbativity of the model parameters provide upper bounds for the derivative coupling. Specifically, for the quadratic and quartic models, these bounds are $|\tilde{\alpha}| \lesssim 10^{19}$ and $|\tilde{\alpha}| \lesssim 10^{15}$, respectively. We have also studied reheating in the simpler case of the quadratic model that allows for approximate analytic treatment and found a maximum reheating temperature of $\mathcal{O}(3\times 10^{15}) \GeV$ for small values of the quadratic Ricci scalar parameter. 

\acknowledgments
%-------------------------------------------------------------------------------

The work of IDG was supported by the Estonian Research Council grants MOBJD1202, RVTT3,  RVTT7, and by the CoE program TK202 ``Fundamental Universe''.

\bibliography{der_coupl}{}

\providecommand{\href}[2]{#2}\begingroup\raggedright\begin{thebibliography}{100}

\bibitem{Kazanas:1980tx}
D.~Kazanas, {\it {Dynamics of the Universe and Spontaneous Symmetry Breaking}},  {\em Astrophys. J. Lett.} {\bf 241} (1980) L59--L63.

\bibitem{Sato:1980yn}
K.~Sato, {\it {First Order Phase Transition of a Vacuum and Expansion of the Universe}},  {\em Mon. Not. Roy. Astron. Soc.} {\bf 195} (1981) 467--479.

\bibitem{Guth:1980zm}
A.~H. Guth, {\it {The Inflationary Universe: A Possible Solution to the Horizon and Flatness Problems}},  {\em Phys. Rev. D} {\bf 23} (1981) 347--356.

\bibitem{Linde:1981mu}
A.~D. Linde, {\it {A New Inflationary Universe Scenario: A Possible Solution of the Horizon, Flatness, Homogeneity, Isotropy and Primordial Monopole Problems}},  {\em Phys. Lett. B} {\bf 108} (1982) 389--393.

\bibitem{Starobinsky:1979ty}
A.~A. Starobinsky, {\it {Spectrum of relict gravitational radiation and the early state of the universe}},  {\em JETP Lett.} {\bf 30} (1979) 682--685.

\bibitem{Mukhanov:1981xt}
V.~F. Mukhanov and G.~V. Chibisov, {\it {Quantum Fluctuations and a Nonsingular Universe}},  {\em JETP Lett.} {\bf 33} (1981) 532--535.

\bibitem{Hawking:1982cz}
S.~W. Hawking, {\it {The Development of Irregularities in a Single Bubble Inflationary Universe}},  {\em Phys. Lett. B} {\bf 115} (1982) 295.

\bibitem{Starobinsky:1982ee}
A.~A. Starobinsky, {\it {Dynamics of Phase Transition in the New Inflationary Universe Scenario and Generation of Perturbations}},  {\em Phys. Lett. B} {\bf 117} (1982) 175--178.

\bibitem{Guth:1982ec}
A.~H. Guth and S.~Y. Pi, {\it {Fluctuations in the New Inflationary Universe}},  {\em Phys. Rev. Lett.} {\bf 49} (1982) 1110--1113.

\bibitem{Bardeen:1983qw}
J.~M. Bardeen, P.~J. Steinhardt, and M.~S. Turner, {\it {Spontaneous Creation of Almost Scale - Free Density Perturbations in an Inflationary Universe}},  {\em Phys. Rev. D} {\bf 28} (1983) 679.

\bibitem{Bezrukov:2007ep}
F.~L. Bezrukov and M.~Shaposhnikov, {\it {The Standard Model Higgs boson as the inflaton}},  {\em Phys. Lett. B} {\bf 659} (2008) 703--706, [\href{http://arxiv.org/abs/0710.3755}{{\tt arXiv:0710.3755}}].

\bibitem{Amendola:1993uh}
L.~Amendola, {\it {Cosmology with nonminimal derivative couplings}},  {\em Phys. Lett. B} {\bf 301} (1993) 175--182, [\href{http://arxiv.org/abs/gr-qc/9302010}{{\tt gr-qc/9302010}}].

\bibitem{Capozziello:1999uwa}
S.~Capozziello and G.~Lambiase, {\it {Nonminimal derivative coupling and the recovering of cosmological constant}},  {\em Gen. Rel. Grav.} {\bf 31} (1999) 1005--1014, [\href{http://arxiv.org/abs/gr-qc/9901051}{{\tt gr-qc/9901051}}].

\bibitem{Capozziello:1999xt}
S.~Capozziello, G.~Lambiase, and H.~J. Schmidt, {\it {Nonminimal derivative couplings and inflation in generalized theories of gravity}},  {\em Annalen Phys.} {\bf 9} (2000) 39--48, [\href{http://arxiv.org/abs/gr-qc/9906051}{{\tt gr-qc/9906051}}].

\bibitem{Germani:2010gm}
C.~Germani and A.~Kehagias, {\it {New Model of Inflation with Non-minimal Derivative Coupling of Standard Model Higgs Boson to Gravity}},  {\em Phys. Rev. Lett.} {\bf 105} (2010) 011302, [\href{http://arxiv.org/abs/1003.2635}{{\tt arXiv:1003.2635}}].

\bibitem{Tsujikawa:2012mk}
S.~Tsujikawa, {\it {Observational tests of inflation with a field derivative coupling to gravity}},  {\em Phys. Rev. D} {\bf 85} (2012) 083518, [\href{http://arxiv.org/abs/1201.5926}{{\tt arXiv:1201.5926}}].

\bibitem{Kamada:2012se}
K.~Kamada, T.~Kobayashi, T.~Takahashi, M.~Yamaguchi, and J.~Yokoyama, {\it {Generalized Higgs inflation}},  {\em Phys. Rev. D} {\bf 86} (2012) 023504, [\href{http://arxiv.org/abs/1203.4059}{{\tt arXiv:1203.4059}}].

\bibitem{Sadjadi:2012zp}
H.~M. Sadjadi and P.~Goodarzi, {\it {Reheating in nonminimal derivative coupling model}},  {\em JCAP} {\bf 02} (2013) 038, [\href{http://arxiv.org/abs/1203.1580}{{\tt arXiv:1203.1580}}].

\bibitem{Koutsoumbas:2013boa}
G.~Koutsoumbas, K.~Ntrekis, and E.~Papantonopoulos, {\it {Gravitational Particle Production in Gravity Theories with Non-minimal Derivative Couplings}},  {\em JCAP} {\bf 08} (2013) 027, [\href{http://arxiv.org/abs/1305.5741}{{\tt arXiv:1305.5741}}].

\bibitem{Ema:2015oaa}
Y.~Ema, R.~Jinno, K.~Mukaida, and K.~Nakayama, {\it {Particle Production after Inflation with Non-minimal Derivative Coupling to Gravity}},  {\em JCAP} {\bf 10} (2015) 020, [\href{http://arxiv.org/abs/1504.07119}{{\tt arXiv:1504.07119}}].

\bibitem{Gumjudpai:2015vio}
B.~Gumjudpai and P.~Rangdee, {\it {Non-minimal derivative coupling gravity in cosmology}},  {\em Gen. Rel. Grav.} {\bf 47} (2015), no.~11 140, [\href{http://arxiv.org/abs/1511.00491}{{\tt arXiv:1511.00491}}].

\bibitem{Zhu:2015lry}
Y.~Zhu and Y.~Gong, {\it {PPN parameters in gravitational theory with nonminimally derivative coupling}},  {\em Int. J. Mod. Phys. D} {\bf 26} (2016), no.~02 1750005, [\href{http://arxiv.org/abs/1512.05555}{{\tt arXiv:1512.05555}}].

\bibitem{Sheikhahmadi:2016wyz}
H.~Sheikhahmadi, E.~N. Saridakis, A.~Aghamohammadi, and K.~Saaidi, {\it {Hamilton-Jacobi formalism for inflation with non-minimal derivative coupling}},  {\em JCAP} {\bf 10} (2016) 021, [\href{http://arxiv.org/abs/1603.03883}{{\tt arXiv:1603.03883}}].

\bibitem{Dalianis:2016wpu}
I.~Dalianis, G.~Koutsoumbas, K.~Ntrekis, and E.~Papantonopoulos, {\it {Reheating predictions in Gravity Theories with Derivative Coupling}},  {\em JCAP} {\bf 02} (2017) 027, [\href{http://arxiv.org/abs/1608.04543}{{\tt arXiv:1608.04543}}].

\bibitem{Harko:2016xip}
T.~Harko, F.~S.~N. Lobo, E.~N. Saridakis, and M.~Tsoukalas, {\it {Cosmological models in modified gravity theories with extended nonminimal derivative couplings}},  {\em Phys. Rev. D} {\bf 95} (2017), no.~4 044019, [\href{http://arxiv.org/abs/1609.01503}{{\tt arXiv:1609.01503}}].

\bibitem{Tumurtushaa:2019bmc}
G.~Tumurtushaa, {\it {Inflation with Derivative Self-interaction and Coupling to Gravity}},  {\em Eur. Phys. J. C} {\bf 79} (2019), no.~11 920, [\href{http://arxiv.org/abs/1903.05354}{{\tt arXiv:1903.05354}}].

\bibitem{Fu:2019ttf}
C.~Fu, P.~Wu, and H.~Yu, {\it {Primordial Black Holes from Inflation with Nonminimal Derivative Coupling}},  {\em Phys. Rev. D} {\bf 100} (2019), no.~6 063532, [\href{http://arxiv.org/abs/1907.05042}{{\tt arXiv:1907.05042}}].

\bibitem{Dalianis:2019vit}
I.~Dalianis, S.~Karydas, and E.~Papantonopoulos, {\it {Generalized Non-Minimal Derivative Coupling: Application to Inflation and Primordial Black Hole Production}},  {\em JCAP} {\bf 06} (2020) 040, [\href{http://arxiv.org/abs/1910.00622}{{\tt arXiv:1910.00622}}].

\bibitem{Sato:2020ghj}
S.~Sato and K.-i. Maeda, {\it {Stability of hybrid Higgs inflation}},  {\em Phys. Rev. D} {\bf 101} (2020), no.~10 103520, [\href{http://arxiv.org/abs/2001.00154}{{\tt arXiv:2001.00154}}].

\bibitem{Karydas:2021wmx}
S.~Karydas, E.~Papantonopoulos, and E.~N. Saridakis, {\it {Successful Higgs inflation from combined nonminimal and derivative couplings}},  {\em Phys. Rev. D} {\bf 104} (2021), no.~2 023530, [\href{http://arxiv.org/abs/2102.08450}{{\tt arXiv:2102.08450}}].

\bibitem{Stelle:1976gc}
K.~S. Stelle, {\it {Renormalization of Higher Derivative Quantum Gravity}},  {\em Phys. Rev. D} {\bf 16} (1977) 953--969.

\bibitem{Starobinsky1980}
A.~A. Starobinsky, {\it {A New Type of Isotropic Cosmological Models Without Singularity}},  {\em Phys. Lett.} {\bf 91B} (1980) 99--102.

\bibitem{Meng:2004yf}
X.-H. Meng and P.~Wang, {\it {$R^2$ corrections to the cosmological dynamics of inflation in the Palatini formulation}},  {\em Class. Quant. Grav.} {\bf 21} (2004) 2029--2036, [\href{http://arxiv.org/abs/gr-qc/0402011}{{\tt gr-qc/0402011}}].

\bibitem{Borunda:2008kf}
M.~Borunda, B.~Janssen, and M.~Bastero-Gil, {\it {Palatini versus metric formulation in higher curvature gravity}},  {\em JCAP} {\bf 11} (2008) 008, [\href{http://arxiv.org/abs/0804.4440}{{\tt arXiv:0804.4440}}].

\bibitem{Bombacigno:2018tyw}
F.~Bombacigno and G.~Montani, {\it {Big bounce cosmology for Palatini $R^2$ gravity with a Nieh\textendash{}Yan term}},  {\em Eur. Phys. J. C} {\bf 79} (2019), no.~5 405, [\href{http://arxiv.org/abs/1809.07563}{{\tt arXiv:1809.07563}}].

\bibitem{Enckell:2018hmo}
V.-M. Enckell, K.~Enqvist, S.~Rasanen, and L.-P. Wahlman, {\it {Inflation with $R^2$ term in the Palatini formalism}},  {\em JCAP} {\bf 02} (2019) 022, [\href{http://arxiv.org/abs/1810.05536}{{\tt arXiv:1810.05536}}].

\bibitem{Iosifidis:2018zjj}
D.~Iosifidis, A.~C. Petkou, and C.~G. Tsagas, {\it {Torsion/non-metricity duality in f(R) gravity}},  {\em Gen. Rel. Grav.} {\bf 51} (2019), no.~5 66, [\href{http://arxiv.org/abs/1810.06602}{{\tt arXiv:1810.06602}}].

\bibitem{Antoniadis:2018ywb}
I.~Antoniadis, A.~Karam, A.~Lykkas, and K.~Tamvakis, {\it {Palatini inflation in models with an $R^2$ term}},  {\em JCAP} {\bf 11} (2018) 028, [\href{http://arxiv.org/abs/1810.10418}{{\tt arXiv:1810.10418}}].

\bibitem{Antoniadis:2018yfq}
I.~Antoniadis, A.~Karam, A.~Lykkas, T.~Pappas, and K.~Tamvakis, {\it {Rescuing Quartic and Natural Inflation in the Palatini Formalism}},  {\em JCAP} {\bf 03} (2019) 005, [\href{http://arxiv.org/abs/1812.00847}{{\tt arXiv:1812.00847}}].

\bibitem{Tenkanen:2019jiq}
T.~Tenkanen, {\it {Minimal Higgs inflation with an $R^2$ term in Palatini gravity}},  {\em Phys. Rev. D} {\bf 99} (2019), no.~6 063528, [\href{http://arxiv.org/abs/1901.01794}{{\tt arXiv:1901.01794}}].

\bibitem{Edery:2019txq}
A.~Edery and Y.~Nakayama, {\it {Palatini formulation of pure $R^2$ gravity yields Einstein gravity with no massless scalar}},  {\em Phys. Rev. D} {\bf 99} (2019), no.~12 124018, [\href{http://arxiv.org/abs/1902.07876}{{\tt arXiv:1902.07876}}].

\bibitem{Giovannini:2019mgk}
M.~Giovannini, {\it {Post-inflationary phases stiffer than radiation and Palatini formulation}},  {\em Class. Quant. Grav.} {\bf 36} (2019), no.~23 235017, [\href{http://arxiv.org/abs/1905.06182}{{\tt arXiv:1905.06182}}].

\bibitem{Gialamas:2019nly}
I.~D. Gialamas and A.~Lahanas, {\it {Reheating in $R^2$ Palatini inflationary models}},  {\em Phys. Rev. D} {\bf 101} (2020), no.~8 084007, [\href{http://arxiv.org/abs/1911.11513}{{\tt arXiv:1911.11513}}].

\bibitem{Lloyd-Stubbs:2020pvx}
A.~Lloyd-Stubbs and J.~McDonald, {\it {Sub-Planckian $\phi^2$ inflation in the Palatini formulation of gravity with an $R^2$ term}},  {\em Phys. Rev. D} {\bf 101} (2020), no.~12 123515, [\href{http://arxiv.org/abs/2002.08324}{{\tt arXiv:2002.08324}}].

\bibitem{Antoniadis:2020dfq}
I.~Antoniadis, A.~Lykkas, and K.~Tamvakis, {\it {Constant-roll in the Palatini-$R^2$ models}},  {\em JCAP} {\bf 04} (2020), no.~04 033, [\href{http://arxiv.org/abs/2002.12681}{{\tt arXiv:2002.12681}}].

\bibitem{Ghilencea:2020piz}
D.~M. Ghilencea, {\it {Palatini quadratic gravity: spontaneous breaking of gauged scale symmetry and inflation}},  {\em Eur. Phys. J. C} {\bf 80} (4, 2020) 1147, [\href{http://arxiv.org/abs/2003.08516}{{\tt arXiv:2003.08516}}].

\bibitem{Das:2020kff}
N.~Das and S.~Panda, {\it {Inflation and Reheating in f(R,h) theory formulated in the Palatini formalism}},  {\em JCAP} {\bf 05} (2021) 019, [\href{http://arxiv.org/abs/2005.14054}{{\tt arXiv:2005.14054}}].

\bibitem{Gialamas:2020snr}
I.~D. Gialamas, A.~Karam, and A.~Racioppi, {\it {Dynamically induced Planck scale and inflation in the Palatini formulation}},  {\em JCAP} {\bf 11} (2020) 014, [\href{http://arxiv.org/abs/2006.09124}{{\tt arXiv:2006.09124}}].

\bibitem{Ghilencea:2020rxc}
D.~M. Ghilencea, {\it {Gauging scale symmetry and inflation: Weyl versus Palatini gravity}},  {\em Eur. Phys. J. C} {\bf 81} (2021), no.~6 510, [\href{http://arxiv.org/abs/2007.14733}{{\tt arXiv:2007.14733}}].

\bibitem{Iosifidis:2020dck}
D.~Iosifidis and L.~Ravera, {\it {Parity Violating Metric-Affine Gravity Theories}},  {\em Class. Quant. Grav.} {\bf 38} (2021), no.~11 115003, [\href{http://arxiv.org/abs/2009.03328}{{\tt arXiv:2009.03328}}].

\bibitem{Bekov:2020dww}
S.~Bekov, K.~Myrzakulov, R.~Myrzakulov, and D.~S.-C. G\'omez, {\it {General slow-roll inflation in $f(R)$ gravity under the Palatini approach}},  {\em Symmetry} {\bf 12} (2020), no.~12 1958, [\href{http://arxiv.org/abs/2010.12360}{{\tt arXiv:2010.12360}}].

\bibitem{Dimopoulos:2020pas}
K.~Dimopoulos and S.~S\'anchez~L\'opez, {\it {Quintessential inflation in Palatini $f(R)$ gravity}},  {\em Phys. Rev. D} {\bf 103} (2021), no.~4 043533, [\href{http://arxiv.org/abs/2012.06831}{{\tt arXiv:2012.06831}}].

\bibitem{Karam:2021sno}
A.~Karam, E.~Tomberg, and H.~Veerm\"ae, {\it {Tachyonic preheating in Palatini $R^2$ inflation}},  {\em JCAP} {\bf 06} (2021) 023, [\href{http://arxiv.org/abs/2102.02712}{{\tt arXiv:2102.02712}}].

\bibitem{Lykkas:2021vax}
A.~Lykkas and K.~Tamvakis, {\it {Extended interactions in the Palatini-$R^2$ inflation}},  {\em JCAP} {\bf 08} (2021), no.~043 [\href{http://arxiv.org/abs/2103.10136}{{\tt arXiv:2103.10136}}].

\bibitem{Gialamas:2021enw}
I.~D. Gialamas, A.~Karam, T.~D. Pappas, and V.~C. Spanos, {\it {Scale-invariant quadratic gravity and inflation in the Palatini formalism}},  {\em Phys. Rev. D} {\bf 104} (2021), no.~2 023521, [\href{http://arxiv.org/abs/2104.04550}{{\tt arXiv:2104.04550}}].

\bibitem{Antoniadis:2021axu}
I.~Antoniadis, A.~Guillen, and K.~Tamvakis, {\it {Ultraviolet behaviour of Higgs inflation models}},  {\em JHEP} {\bf 08} (2021) 018, [\href{http://arxiv.org/abs/2106.09390}{{\tt arXiv:2106.09390}}]. [Addendum: JHEP 05, 074 (2022)].

\bibitem{Gialamas:2021rpr}
I.~D. Gialamas, A.~Karam, T.~D. Pappas, A.~Racioppi, and V.~C. Spanos, {\it {Scale-invariance, dynamically induced Planck scale and inflation in the Palatini formulation}},  {\em J. Phys. Conf. Ser.} {\bf 2105} (2021), no.~1 012005, [\href{http://arxiv.org/abs/2107.04408}{{\tt arXiv:2107.04408}}].

\bibitem{AlHallak:2021hwb}
M.~AlHallak, A.~AlRakik, N.~Chamoun, and M.~S. El-Daher, {\it {Palatini f(R) Gravity and Variants of k-/Constant Roll/Warm Inflation within Variation of Strong Coupling Scenario}},  {\em Universe} {\bf 8} (2022), no.~2 126, [\href{http://arxiv.org/abs/2111.05075}{{\tt arXiv:2111.05075}}].

\bibitem{Dioguardi:2021fmr}
C.~Dioguardi, A.~Racioppi, and E.~Tomberg, {\it {Slow-roll inflation in Palatini F(R) gravity}},  {\em JHEP} {\bf 06} (2022) 106, [\href{http://arxiv.org/abs/2112.12149}{{\tt arXiv:2112.12149}}].

\bibitem{Dimopoulos:2022tvn}
K.~Dimopoulos, A.~Karam, S.~S\'anchez~L\'opez, and E.~Tomberg, {\it {Modelling Quintessential Inflation in Palatini-Modified Gravity}},  {\em Galaxies} {\bf 10} (2022), no.~2 57, [\href{http://arxiv.org/abs/2203.05424}{{\tt arXiv:2203.05424}}].

\bibitem{Dimopoulos:2022rdp}
K.~Dimopoulos, A.~Karam, S.~S\'anchez~L\'opez, and E.~Tomberg, {\it {Palatini R $^{2}$ quintessential inflation}},  {\em JCAP} {\bf 10} (2022) 076, [\href{http://arxiv.org/abs/2206.14117}{{\tt arXiv:2206.14117}}].

\bibitem{Pradisi:2022nmh}
G.~Pradisi and A.~Salvio, {\it {(In)equivalence of metric-affine and metric effective field theories}},  {\em Eur. Phys. J. C} {\bf 82} (2022), no.~9 840, [\href{http://arxiv.org/abs/2206.15041}{{\tt arXiv:2206.15041}}].

\bibitem{Durrer:2022emo}
R.~Durrer, O.~Sobol, and S.~Vilchinskii, {\it {Magnetogenesis in Higgs-Starobinsky inflation}},  {\em Phys. Rev. D} {\bf 106} (2022), no.~12 123520, [\href{http://arxiv.org/abs/2207.05030}{{\tt arXiv:2207.05030}}].

\bibitem{Salvio:2022suk}
A.~Salvio, {\it {Inflating and reheating the Universe with an independent affine connection}},  {\em Phys. Rev. D} {\bf 106} (2022), no.~10 103510, [\href{http://arxiv.org/abs/2207.08830}{{\tt arXiv:2207.08830}}].

\bibitem{Antoniadis:2022cqh}
I.~Antoniadis, A.~Guillen, and K.~Tamvakis, {\it {Late time acceleration in Palatini gravity}},  {\em JHEP} {\bf 11} (2022) 144, [\href{http://arxiv.org/abs/2207.13732}{{\tt arXiv:2207.13732}}].

\bibitem{Lahanas:2022mng}
A.~B. Lahanas, {\it {Issues in Palatini $R^2$ inflation: Bounds on the reheating temperature}},  {\em Phys. Rev. D} {\bf 106} (2022), no.~12 123530, [\href{http://arxiv.org/abs/2210.00837}{{\tt arXiv:2210.00837}}].

\bibitem{Gialamas:2022xtt}
I.~D. Gialamas and K.~Tamvakis, {\it {Inflation in metric-affine quadratic gravity}},  {\em JCAP} {\bf 03} (2023) 042, [\href{http://arxiv.org/abs/2212.09896}{{\tt arXiv:2212.09896}}].

\bibitem{Dioguardi:2022oqu}
C.~Dioguardi, A.~Racioppi, and E.~Tomberg, {\it {Inflation in Palatini quadratic gravity (and beyond)}},  \href{http://arxiv.org/abs/2212.11869}{{\tt arXiv:2212.11869}}.

\bibitem{Iosifidis:2022xvp}
D.~Iosifidis, R.~Myrzakulov, and L.~Ravera, {\it {Cosmology of Metric-Affine R+\ensuremath{\beta}$R^2$ Gravity with Pure Shear Hypermomentum}},  {\em Fortsch. Phys.} {\bf 72} (2024), no.~1 2300003, [\href{http://arxiv.org/abs/2301.00669}{{\tt arXiv:2301.00669}}].

\bibitem{Gialamas:2023aim}
I.~D. Gialamas and K.~Tamvakis, {\it {Bimetric-affine quadratic gravity}},  {\em Phys. Rev. D} {\bf 107} (2023), no.~10 104012, [\href{http://arxiv.org/abs/2303.11353}{{\tt arXiv:2303.11353}}].

\bibitem{Gialamas:2023flv}
I.~D. Gialamas, A.~Karam, T.~D. Pappas, and E.~Tomberg, {\it {Implications of Palatini gravity for inflation and beyond}},  \href{http://arxiv.org/abs/2303.14148}{{\tt arXiv:2303.14148}}.

\bibitem{SanchezLopez:2023ixx}
S.~S\'anchez~L\'opez, K.~Dimopoulos, A.~Karam, and E.~Tomberg, {\it {Observable gravitational waves from hyperkination in Palatini gravity and beyond}},  {\em Eur. Phys. J. C} {\bf 83} (2023), no.~12 1152, [\href{http://arxiv.org/abs/2305.01399}{{\tt arXiv:2305.01399}}].

\bibitem{Dioguardi:2023jwa}
C.~Dioguardi and A.~Racioppi, {\it {Palatini $F(R,X)$: a new framework for inflationary attractors}},  \href{http://arxiv.org/abs/2307.02963}{{\tt arXiv:2307.02963}}.

\bibitem{DiMarco:2023ncs}
A.~Di~Marco, E.~Orazi, and G.~Pradisi, {\it {Einstein\textendash{}Cartan pseudoscalaron inflation}},  {\em Eur. Phys. J. C} {\bf 84} (2024), no.~2 146, [\href{http://arxiv.org/abs/2309.11345}{{\tt arXiv:2309.11345}}].

\bibitem{Gomes:2023xzk}
D.~A. Gomes, R.~Briffa, A.~Kozak, J.~Levi~Said, M.~Saal, and A.~Wojnar, {\it {Cosmological constraints of Palatini f(R) gravity}},  {\em JCAP} {\bf 01} (2024) 011, [\href{http://arxiv.org/abs/2310.17339}{{\tt arXiv:2310.17339}}].

\bibitem{Hu:2023yjn}
W.-Y. Hu, Q.-Y. Wang, Y.-Q. Ma, and Y.~Tang, {\it {Gravitational Waves from Preheating in Inflation with Weyl Symmetry}},  \href{http://arxiv.org/abs/2311.00239}{{\tt arXiv:2311.00239}}.

\bibitem{Bauer:2008zj}
F.~Bauer and D.~A. Demir, {\it {Inflation with Non-Minimal Coupling: Metric versus Palatini Formulations}},  {\em Phys. Lett.} {\bf B665} (2008) 222--226, [\href{http://arxiv.org/abs/0803.2664}{{\tt arXiv:0803.2664}}].

\bibitem{Rasanen:2017ivk}
S.~Rasanen and P.~Wahlman, {\it {Higgs inflation with loop corrections in the Palatini formulation}},  {\em JCAP} {\bf 11} (2017) 047, [\href{http://arxiv.org/abs/1709.07853}{{\tt arXiv:1709.07853}}].

\bibitem{Tenkanen:2017jih}
T.~Tenkanen, {\it {Resurrecting Quadratic Inflation with a non-minimal coupling to gravity}},  {\em JCAP} {\bf 12} (2017) 001, [\href{http://arxiv.org/abs/1710.02758}{{\tt arXiv:1710.02758}}].

\bibitem{Racioppi:2017spw}
A.~Racioppi, {\it {Coleman-Weinberg linear inflation: metric vs. Palatini formulation}},  {\em JCAP} {\bf 12} (2017) 041, [\href{http://arxiv.org/abs/1710.04853}{{\tt arXiv:1710.04853}}].

\bibitem{Markkanen:2017tun}
T.~Markkanen, T.~Tenkanen, V.~Vaskonen, and H.~Veerm\"ae, {\it {Quantum corrections to quartic inflation with a non-minimal coupling: metric vs. Palatini}},  {\em JCAP} {\bf 03} (2018) 029, [\href{http://arxiv.org/abs/1712.04874}{{\tt arXiv:1712.04874}}].

\bibitem{Jarv:2017azx}
L.~J\"arv, A.~Racioppi, and T.~Tenkanen, {\it {Palatini side of inflationary attractors}},  {\em Phys. Rev. D} {\bf 97} (2018), no.~8 083513, [\href{http://arxiv.org/abs/1712.08471}{{\tt arXiv:1712.08471}}].

\bibitem{Fu:2017iqg}
C.~Fu, P.~Wu, and H.~Yu, {\it {Inflationary dynamics and preheating of the nonminimally coupled inflaton field in the metric and Palatini formalisms}},  {\em Phys. Rev. D} {\bf 96} (2017), no.~10 103542, [\href{http://arxiv.org/abs/1801.04089}{{\tt arXiv:1801.04089}}].

\bibitem{Racioppi:2018zoy}
A.~Racioppi, {\it {New universal attractor in nonminimally coupled gravity: Linear inflation}},  {\em Phys. Rev. D} {\bf 97} (2018), no.~12 123514, [\href{http://arxiv.org/abs/1801.08810}{{\tt arXiv:1801.08810}}].

\bibitem{Kozak:2018vlp}
A.~Kozak and A.~Borowiec, {\it {Palatini frames in scalar\textendash{}tensor theories of gravity}},  {\em Eur. Phys. J. C} {\bf 79} (2019), no.~4 335, [\href{http://arxiv.org/abs/1808.05598}{{\tt arXiv:1808.05598}}].

\bibitem{Rasanen:2018ihz}
S.~Rasanen, {\it {Higgs inflation in the Palatini formulation with kinetic terms for the metric}},  {\em Open J. Astrophys.} {\bf 2} (2019), no.~1 1, [\href{http://arxiv.org/abs/1811.09514}{{\tt arXiv:1811.09514}}].

\bibitem{Almeida:2018oid}
J.~P.~B. Almeida, N.~Bernal, J.~Rubio, and T.~Tenkanen, {\it {Hidden Inflaton Dark Matter}},  {\em JCAP} {\bf 03} (2019) 012, [\href{http://arxiv.org/abs/1811.09640}{{\tt arXiv:1811.09640}}].

\bibitem{Shimada:2018lnm}
K.~Shimada, K.~Aoki, and K.-i. Maeda, {\it {Metric-affine Gravity and Inflation}},  {\em Phys. Rev. D} {\bf 99} (2019), no.~10 104020, [\href{http://arxiv.org/abs/1812.03420}{{\tt arXiv:1812.03420}}].

\bibitem{Takahashi:2018brt}
T.~Takahashi and T.~Tenkanen, {\it {Towards distinguishing variants of non-minimal inflation}},  {\em JCAP} {\bf 04} (2019) 035, [\href{http://arxiv.org/abs/1812.08492}{{\tt arXiv:1812.08492}}].

\bibitem{Jinno:2018jei}
R.~Jinno, K.~Kaneta, K.-y. Oda, and S.~C. Park, {\it {Hillclimbing inflation in metric and Palatini formulations}},  {\em Phys. Lett. B} {\bf 791} (2019) 396--402, [\href{http://arxiv.org/abs/1812.11077}{{\tt arXiv:1812.11077}}].

\bibitem{Rubio:2019ypq}
J.~Rubio and E.~S. Tomberg, {\it {Preheating in Palatini Higgs inflation}},  {\em JCAP} {\bf 1904} (2019), no.~04 021, [\href{http://arxiv.org/abs/1902.10148}{{\tt arXiv:1902.10148}}].

\bibitem{Racioppi:2019jsp}
A.~Racioppi, {\it {Non-Minimal (Self-)Running Inflation: Metric vs. Palatini Formulation}},  {\em JHEP} {\bf 21} (2020) 011, [\href{http://arxiv.org/abs/1912.10038}{{\tt arXiv:1912.10038}}].

\bibitem{Shaposhnikov:2020fdv}
M.~Shaposhnikov, A.~Shkerin, and S.~Zell, {\it {Quantum Effects in Palatini Higgs Inflation}},  {\em JCAP} {\bf 07} (2020) 064, [\href{http://arxiv.org/abs/2002.07105}{{\tt arXiv:2002.07105}}].

\bibitem{Borowiec:2020lfx}
A.~Borowiec and A.~Kozak, {\it {New class of hybrid metric-Palatini scalar-tensor theories of gravity}},  {\em JCAP} {\bf 07} (2020) 003, [\href{http://arxiv.org/abs/2003.02741}{{\tt arXiv:2003.02741}}].

\bibitem{Jarv:2020qqm}
L.~J\"arv, A.~Karam, A.~Kozak, A.~Lykkas, A.~Racioppi, and M.~Saal, {\it {Equivalence of inflationary models between the metric and Palatini formulation of scalar-tensor theories}},  {\em Phys. Rev. D} {\bf 102} (2020), no.~4 044029, [\href{http://arxiv.org/abs/2005.14571}{{\tt arXiv:2005.14571}}].

\bibitem{Karam:2020rpa}
A.~Karam, M.~Raidal, and E.~Tomberg, {\it {Gravitational dark matter production in Palatini preheating}},  {\em JCAP} {\bf 03} (2021) 064, [\href{http://arxiv.org/abs/2007.03484}{{\tt arXiv:2007.03484}}].

\bibitem{McDonald:2020lpz}
J.~McDonald, {\it {Does Palatini Higgs Inflation Conserve Unitarity?}},  {\em JCAP} {\bf 04} (2021) 069, [\href{http://arxiv.org/abs/2007.04111}{{\tt arXiv:2007.04111}}].

\bibitem{Langvik:2020nrs}
M.~Langvik, J.-M. Ojanper\"a, S.~Raatikainen, and S.~Rasanen, {\it {Higgs inflation with the Holst and the Nieh\textendash{}Yan term}},  {\em Phys. Rev. D} {\bf 103} (2021), no.~8 083514, [\href{http://arxiv.org/abs/2007.12595}{{\tt arXiv:2007.12595}}].

\bibitem{Shaposhnikov:2020gts}
M.~Shaposhnikov, A.~Shkerin, I.~Timiryasov, and S.~Zell, {\it {Higgs inflation in Einstein-Cartan gravity}},  {\em JCAP} {\bf 02} (2021) 008, [\href{http://arxiv.org/abs/2007.14978}{{\tt arXiv:2007.14978}}]. [Erratum: JCAP 10, E01 (2021)].

\bibitem{Shaposhnikov:2020frq}
M.~Shaposhnikov, A.~Shkerin, I.~Timiryasov, and S.~Zell, {\it {Einstein-Cartan gravity, matter, and scale-invariant generalization}},  {\em JHEP} {\bf 10} (2020) 177, [\href{http://arxiv.org/abs/2007.16158}{{\tt arXiv:2007.16158}}].

\bibitem{Mikura:2020qhc}
Y.~Mikura, Y.~Tada, and S.~Yokoyama, {\it {Conformal inflation in the metric-affine geometry}},  {\em EPL} {\bf 132} (2020), no.~3 39001, [\href{http://arxiv.org/abs/2008.00628}{{\tt arXiv:2008.00628}}].

\bibitem{Verner:2020gfa}
S.~Verner, {\it {Quintessential Inflation in Palatini Gravity}},  {\em JCAP} {\bf 04} (2021) [\href{http://arxiv.org/abs/2010.11201}{{\tt arXiv:2010.11201}}].

\bibitem{Enckell:2020lvn}
V.-M. Enckell, S.~Nurmi, S.~R\"as\"anen, and E.~Tomberg, {\it {Critical point Higgs inflation in the Palatini formulation}},  {\em JHEP} {\bf 04} (2021) 059, [\href{http://arxiv.org/abs/2012.03660}{{\tt arXiv:2012.03660}}].

\bibitem{Reyimuaji:2020goi}
Y.~Reyimuaji and X.~Zhang, {\it {Natural inflation with a nonminimal coupling to gravity}},  {\em JCAP} {\bf 03} (2021) 059, [\href{http://arxiv.org/abs/2012.14248}{{\tt arXiv:2012.14248}}].

\bibitem{Karam:2021wzz}
A.~Karam, S.~Karamitsos, and M.~Saal, {\it {\ensuremath{\beta}-function reconstruction of Palatini inflationary attractors}},  {\em JCAP} {\bf 10} (2021) 068, [\href{http://arxiv.org/abs/2103.01182}{{\tt arXiv:2103.01182}}].

\bibitem{Mikura:2021ldx}
Y.~Mikura, Y.~Tada, and S.~Yokoyama, {\it {Minimal $k$-inflation in light of the conformal metric-affine geometry}},  {\em Phys. Rev. D} {\bf 103} (2021), no.~10 L101303, [\href{http://arxiv.org/abs/2103.13045}{{\tt arXiv:2103.13045}}].

\bibitem{Racioppi:2021ynx}
A.~Racioppi, J.~Rajasalu, and K.~Selke, {\it {Multiple point criticality principle and Coleman-Weinberg inflation}},  {\em JHEP} {\bf 06} (2022) 107, [\href{http://arxiv.org/abs/2109.03238}{{\tt arXiv:2109.03238}}].

\bibitem{Mikura:2021clt}
Y.~Mikura and Y.~Tada, {\it {On UV-completion of Palatini-Higgs inflation}},  {\em JCAP} {\bf 05} (2022), no.~05 035, [\href{http://arxiv.org/abs/2110.03925}{{\tt arXiv:2110.03925}}].

\bibitem{Cheong:2021kyc}
D.~Y. Cheong, S.~M. Lee, and S.~C. Park, {\it {Reheating in models with non-minimal coupling in metric and~Palatini formalisms}},  {\em JCAP} {\bf 02} (2022), no.~02 029, [\href{http://arxiv.org/abs/2111.00825}{{\tt arXiv:2111.00825}}].

\bibitem{Azri:2021uat}
H.~Azri, I.~Bamwidhi, and S.~Nasri, {\it {Isocurvature modes and non-Gaussianity in affine inflation}},  {\em Phys. Rev. D} {\bf 104} (2021), no.~10 104064, [\href{http://arxiv.org/abs/2111.03828}{{\tt arXiv:2111.03828}}].

\bibitem{Racioppi:2021jai}
A.~Racioppi and M.~Vasar, {\it {On the number of e-folds in the Jordan and Einstein frames}},  {\em Eur. Phys. J. Plus} {\bf 137} (2022), no.~5 637, [\href{http://arxiv.org/abs/2111.09677}{{\tt arXiv:2111.09677}}].

\bibitem{Piani:2022gon}
M.~Piani and J.~Rubio, {\it {Higgs-Dilaton inflation in Einstein-Cartan gravity}},  {\em JCAP} {\bf 05} (2022), no.~05 009, [\href{http://arxiv.org/abs/2202.04665}{{\tt arXiv:2202.04665}}].

\bibitem{Karananas:2022byw}
G.~K. Karananas, M.~Shaposhnikov, and S.~Zell, {\it {Field redefinitions, perturbative unitarity and Higgs inflation}},  {\em JHEP} {\bf 06} (2022) 132, [\href{http://arxiv.org/abs/2203.09534}{{\tt arXiv:2203.09534}}].

\bibitem{Rigouzzo:2022yan}
C.~Rigouzzo and S.~Zell, {\it {Coupling metric-affine gravity to a Higgs-like scalar field}},  {\em Phys. Rev. D} {\bf 106} (2022), no.~2 024015, [\href{http://arxiv.org/abs/2204.03003}{{\tt arXiv:2204.03003}}].

\bibitem{Gialamas:2022gxv}
I.~D. Gialamas, A.~Karam, and T.~D. Pappas, {\it {Gravitational corrections to electroweak vacuum decay: metric vs. Palatini}},  {\em Phys. Lett. B} {\bf 840} (2023) 137885, [\href{http://arxiv.org/abs/2212.03052}{{\tt arXiv:2212.03052}}].

\bibitem{Hyun:2023bkf}
S.~C. Hyun, J.~Kim, T.~Kodama, S.~C. Park, and T.~Takahashi, {\it {Nonminimally assisted inflation: a~general analysis}},  {\em JCAP} {\bf 05} (2023) 050, [\href{http://arxiv.org/abs/2302.05866}{{\tt arXiv:2302.05866}}].

\bibitem{Piani:2023aof}
M.~Piani and J.~Rubio, {\it {Preheating in Einstein-Cartan Higgs Inflation: oscillon formation}},  {\em JCAP} {\bf 12} (2023) 002, [\href{http://arxiv.org/abs/2304.13056}{{\tt arXiv:2304.13056}}].

\bibitem{Gialamas:2023emn}
I.~D. Gialamas and H.~Veerm\"ae, {\it {Electroweak vacuum decay in metric-affine gravity}},  {\em Phys. Lett. B} {\bf 844} (2023) 138109, [\href{http://arxiv.org/abs/2305.07693}{{\tt arXiv:2305.07693}}].

\bibitem{Rigouzzo:2023sbb}
C.~Rigouzzo and S.~Zell, {\it {Coupling metric-affine gravity to the standard model and dark matter fermions}},  {\em Phys. Rev. D} {\bf 108} (2023), no.~12 124067, [\href{http://arxiv.org/abs/2306.13134}{{\tt arXiv:2306.13134}}].

\bibitem{Barman:2023opy}
B.~Barman, N.~Bernal, and J.~Rubio, {\it {Rescuing Gravitational-Reheating in Chaotic Inflation}},  \href{http://arxiv.org/abs/2310.06039}{{\tt arXiv:2310.06039}}.

\bibitem{Gialamas:2020vto}
I.~D. Gialamas, A.~Karam, A.~Lykkas, and T.~D. Pappas, {\it {Palatini-Higgs inflation with nonminimal derivative coupling}},  {\em Phys. Rev. D} {\bf 102} (2020), no.~6 063522, [\href{http://arxiv.org/abs/2008.06371}{{\tt arXiv:2008.06371}}].

\bibitem{Nezhad:2023dys}
H.~B. Nezhad and S.~Rasanen, {\it {Scalar fields with derivative coupling to curvature in the Palatini and the metric formulation}},  {\em JCAP} {\bf 02} (2024) 009, [\href{http://arxiv.org/abs/2307.04618}{{\tt arXiv:2307.04618}}].

\bibitem{Annala:2020cqj}
J.~Annala, {\it {Higgs inflation and higher-order gravity in Palatini formulation}},  Master's thesis, Helsinki U., 2020.

\bibitem{Annala:2021zdt}
J.~Annala and S.~Rasanen, {\it {Inflation with $R_{(\ensuremath{\alpha}\ensuremath{\beta})}$ terms in the Palatini formulation}},  {\em JCAP} {\bf 09} (2021) 032, [\href{http://arxiv.org/abs/2106.12422}{{\tt arXiv:2106.12422}}].

\bibitem{BeltranJimenez:2019acz}
J.~Beltr\'an~Jim\'enez and A.~Delhom, {\it {Ghosts in metric-affine higher order curvature gravity}},  {\em Eur. Phys. J. C} {\bf 79} (2019), no.~8 656, [\href{http://arxiv.org/abs/1901.08988}{{\tt arXiv:1901.08988}}].

\bibitem{BeltranJimenez:2020sqf}
J.~Beltr\'an~Jim\'enez and A.~Delhom, {\it {Instabilities in metric-affine theories of gravity with higher order curvature terms}},  {\em Eur. Phys. J. C} {\bf 80} (2020), no.~6 585, [\href{http://arxiv.org/abs/2004.11357}{{\tt arXiv:2004.11357}}].

\bibitem{Marzo:2021iok}
C.~Marzo, {\it {Radiatively stable ghost and tachyon freedom in metric affine gravity}},  {\em Phys. Rev. D} {\bf 106} (2022), no.~2 024045, [\href{http://arxiv.org/abs/2110.14788}{{\tt arXiv:2110.14788}}].

\bibitem{Annala:2022gtl}
J.~Annala and S.~Rasanen, {\it {Stability of non-degenerate Ricci-type Palatini theories}},  {\em JCAP} {\bf 04} (2023) 014, [\href{http://arxiv.org/abs/2212.09820}{{\tt arXiv:2212.09820}}]. [Erratum: JCAP 08, E02 (2023)].

\bibitem{Barker:2024ydb}
W.~Barker and C.~Marzo, {\it {Particle spectra of general Ricci-type Palatini or metric-affine theories}},  \href{http://arxiv.org/abs/2402.07641}{{\tt arXiv:2402.07641}}.

\bibitem{Barker:2024dhb}
W.~Barker and S.~Zell, {\it {Consistent particle physics in metric-affine gravity from extended projective symmetry}},  \href{http://arxiv.org/abs/2402.14917}{{\tt arXiv:2402.14917}}.

\bibitem{Mavromatos:2023wkk}
N.~E. Mavromatos, P.~Pais, and A.~Iorio, {\it {Torsion at Different Scales: From Materials to the Universe}},  {\em Universe} {\bf 9} (2023), no.~12 516, [\href{http://arxiv.org/abs/2310.13150}{{\tt arXiv:2310.13150}}].

\bibitem{Planck:2018jri}
{\bf Planck} Collaboration, Y.~Akrami et~al., {\it {Planck 2018 results. X. Constraints on inflation}},  {\em Astron. Astrophys.} {\bf 641} (2020) A10, [\href{http://arxiv.org/abs/1807.06211}{{\tt arXiv:1807.06211}}].

\bibitem{BICEP:2021xfz}
{\bf BICEP, Keck} Collaboration, P.~A.~R. Ade et~al., {\it {Improved Constraints on Primordial Gravitational Waves using Planck, WMAP, and BICEP/Keck Observations through the 2018 Observing Season}},  {\em Phys. Rev. Lett.} {\bf 127} (2021), no.~15 151301, [\href{http://arxiv.org/abs/2110.00483}{{\tt arXiv:2110.00483}}].

\bibitem{Tristram:2021tvh}
M.~Tristram et~al., {\it {Improved limits on the tensor-to-scalar ratio using BICEP and Planck data}},  {\em Phys. Rev. D} {\bf 105} (2022), no.~8 083524, [\href{http://arxiv.org/abs/2112.07961}{{\tt arXiv:2112.07961}}].

\bibitem{Galloni:2022mok}
G.~Galloni, N.~Bartolo, S.~Matarrese, M.~Migliaccio, A.~Ricciardone, and N.~Vittorio, {\it {Updated constraints on amplitude and tilt of the tensor primordial spectrum}},  {\em JCAP} {\bf 04} (2023) 062, [\href{http://arxiv.org/abs/2208.00188}{{\tt arXiv:2208.00188}}].

\bibitem{Zhang:2023hfx}
F.-Y. Zhang, {\it {Reheating predictions in non-minimally coupled inflationary models with radiative corrections}},  {\em Phys. Dark Univ.} {\bf 39} (2023) 101169.

\end{thebibliography}\endgroup
\end{document}